%% file: p.tex
\documentclass[conference]{IEEEtran}

\usepackage[square,comma,numbers,sort&compress]{natbib}

\include{pkgs}

\newcommand{\sys}{\mbox{\textsc{ArcHeap}}\xspace}

\newcommand{\XXX}[1]{\textcolor{red}{XXX: #1}}

\input{cmds}
\input{rev}

\begin{document}

\input{hdr}
\date{}
\maketitle

\sloppy

\input{abstract}
\input{intro}
\input{background}

\input{challenge}

\input{design}
\input{impl}
\input{app}
\input{eval}
\input{discuss}

\input{relwk}

\input{conclusion}

\clearpage
\bibliographystyle{abbrvnat}
\footnotesize
\setlength{\bibsep}{3pt}
\bibliography{p,sslab,conf}

\clearpage
\nobalance
\input{appendix}

\end{document}

%% file: pkgs.tex
\input{warning}

\usepackage[hyphens]{url}
\usepackage[breaklinks,colorlinks]{hyperref}
\usepackage[usenames,dvipsnames]{xcolor}
\hypersetup{citecolor=blue,linkcolor=blue}
\usepackage{amsmath,amsopn,amssymb}
\usepackage{subcaption}
\usepackage{endnotes,microtype,xspace,graphicx,fancyvrb,multirow}
\usepackage{booktabs}
\usepackage{array,underscore,relsize}
\usepackage[T1]{fontenc}
\usepackage{fancyhdr}
\usepackage{enumitem}
\usepackage[labelfont=bf,font=small,skip=5pt]{caption}
\usepackage{fp}
\usepackage{siunitx}

\usepackage{balance}

\usepackage{multicol}

\usepackage{makecell}
\usepackage[linesnumbered,ruled]{algorithm2e}
\SetAlgorithmName{Algorithm}{Algorithm}{}

\usepackage{graphicx,verbatimbox}
\usepackage{tikz}

\usepackage{amssymb}

\newcommand{\bm}{\vspace{-13px}}
\renewcommand{\arraystretch}{.93}

\usepackage{tabu}
\usepackage{colortbl}

%% file: warning.tex
\usepackage{silence}

%% file: cmds.tex
\newcommand{\cc}[1]{\mbox{\smaller[0.5]\texttt{#1}}}

\fvset{fontsize=\scriptsize,xleftmargin=8pt,numbers=left,numbersep=5pt}

\input{code/fmt}
\newcommand{\figrule}{\hrule width \hsize height .33pt}
\newcommand{\coderule}{\vspace{0.3em}\figrule\vspace{0.3em}}

\def\Snospace~{\S{}}

\newcommand{\x}{$\times$\xspace}

\newif\ifdraft\drafttrue
\newif\ifnotes\notestrue
\ifdraft\else\notesfalse\fi

\input{glyphtounicode}
\newcolumntype{R}[1]{>{\raggedleft\let\newline\\\arraybackslash\hspace{0pt}}p{#1}}

\newcommand{\includepdf}[1]{
  \includegraphics[width=\columnwidth]{#1}
}

\newcommand{\squishlist}{
\vspace{3pt}
\begin{itemize}[noitemsep,nolistsep]
  \setlength{\itemsep}{-0pt}
}
\newcommand{\squishend}{
  \end{itemize}
  \vspace{3pt}
}

\newcommand*\CC[1]{%
\begin{tikzpicture}[baseline=(C.base)]
\node[draw,circle,inner sep=0.2pt](C) {#1};
\end{tikzpicture}}

\newcommand*\BC[1]{%
\begin{tikzpicture}[baseline=(C.base)]
\node[draw,circle,fill=black,inner sep=0.2pt](C) {\textcolor{white}{#1}};
\end{tikzpicture}}

\usepackage{xstring}
\newcommand{\PP}[1]{
\vspace{2px}
\noindent{\bf \IfEndWith{#1}{.}{#1}{#1.}}
}

\newcommand{\V}{\checkmark}

\newcommand{\ptmalloc}{ptmalloc\xspace}
\newcommand{\jemalloc}{jemalloc\xspace}
\newcommand{\tcmalloc}{tcmalloc\xspace}
\newcommand{\dlmalloc}{dlmalloc\xspace}
\newcommand{\PartitionAlloc}{PartitionAlloc\xspace}
\newcommand{\libumem}{libumem\xspace}

\newcommand{\eg}{e.g.,\xspace}
\newcommand{\ie}{i.e.,\xspace}

\newcommand{\malloc}{\cc{malloc}\xspace}
\newcommand{\free}{\cc{free}\xspace}
\newcommand{\fmalloc}{\cc{malloc()}\xspace}
\newcommand{\ffree}{\cc{free()}\xspace}

\newcommand{\etal}{\textit{et al.}}
\newcommand{\consolidate}{\cc{malloc\_consolidate()}\xspace}
\newcommand{\previnuse}{\cc{PREV\_IN\_USE}\xspace}

\newcommand{\precise}{\cc{precise}\xspace}
\newcommand{\trusty}{\cc{trusty}\xspace}
\newcommand{\xenial}{\cc{xenial}\xspace}
\newcommand{\bionic}{\cc{bionic}\xspace}

\newcommand{\reducedactions}{84.3\%\xspace}

\newcommand{\heaphopper}{HeapHopper\xspace}

\newcommand{\newtech}{five\xspace}

\newcommand{\fastbindup}{fast-bin-dup\xspace}

\newcommand{\expsizeactions}{\cc{+Size, TxnList}\xspace}
\newcommand{\expbase}{\cc{Bug+Impact+Chunks}\xspace}
\newcommand{\expsize}{\cc{+Size}\xspace}
\newcommand{\expactions}{\cc{+TxnList}\xspace}

%% file: rev.tex
\gdef\therev{58c7047}
\gdef\thedate{2019-03-01 14:05:17 -0500}

%% file: hdr.tex
\title{Automatic Techniques to Systematically Discover \\
New Heap Exploitation Primitives}

\ifdefined\DRAFT
 \pagestyle{fancyplain}
 \lhead{Rev.~\therev}
 \rhead{\thedate}
 \cfoot{\thepage\ of \pageref{LastPage}}
\fi

\author{
{\rm Insu Yun \quad
     Dhaval Kapil \quad
     Taesoo Kim \quad
} \\
{\it Georgia Institute of Technology}
}

%

%% file: abstract.tex
\begin{abstract}
  Exploitation techniques
  to abuse the metadata of heap allocators
  have been widely studied
  because of their generality (i.e., application independent)
  and powerful capability
  (i.e., bypassing mitigation).
  However, such techniques
  are commonly considered \emph{arts},
  and thus the approaches to discover them
  remain ad-hoc, manual, and allocator-specific at best.

  In this paper,
  we present an automatic tool,
  \sys,
  to systematically discover
  the unexplored heap exploitation primitives,
  regardless of their underlying implementations.
  The key idea of \sys
  is to let the computer autonomously explore the spaces,
  similar in concept to fuzzing,
  by specifying a set of common designs of modern heap allocators
  and root causes of vulnerabilities as models,
  and by providing
  heap operations and attack capabilities
  as actions.
  During the exploration,
  \sys checks
  whether the combinations of these actions
  can be potentially used
  to construct exploitation primitives,
  such as arbitrary write or overlapped chunks.
  As a proof,
  \sys generates working PoC
  that demonstrates
  the discovered exploitation technique.

  We evaluated \sys with three real-world allocators
  (i.e., \ptmalloc, \tcmalloc, and \jemalloc),
  as well as custom allocators
  from the DARPA Cyber Grand Challenge.
  As a result,
  \sys discovered \emph{\newtech previously unknown}
  exploitation primitives in \ptmalloc
  and found several exploitation techniques
  against \jemalloc, \tcmalloc, and even
  custom heap allocators.
  To show the effectiveness of \sys's approach in other domains,
  we also studied how security features evolve
  and which exploit primitives are effective
  across different versions of \ptmalloc.
  
\end{abstract}

%% file: intro.tex
\section{Introduction}
\label{s:intro}


Heap-related vulnerabilities
have been the most common, yet critical source
of security problems
in systems software%
~\cite{weston:windows, lee:dangnull, silvestro:guarder,
silvestro:freeguard}.
According to Microsoft,
heap vulnerabilities
accounted for 53\% of security problems
in their products in 2017~\cite{microsoft-bugs}.
There are two properties
that make heap vulnerabilities
a preferable target for attacks.
First,
heap exploitation techniques
tend to be application-independent,
making it possible to write attack code
without a deep understanding of application internals.
Second,
heap vulnerabilities are typically so powerful
that attackers can easily bypass
modern mitigation schemes
by abusing them.
For example,
a seemingly benign bug
that overwrites \emph{one NULL byte}
to the metadata of \ptmalloc
leads to a privilege escalation
on Chrome OS~\cite{gpz:chrome-os}.

%
%


\begin{table}[!t]
  \centering
  \input{fig/tbl-timeline.tex}
  \caption{
    Timeline for new heap exploitation techniques discovered
    and their count in parentheses (\eg \sys found five new techniques in 2018).
  }
  \bm
  \label{f:tbl-timeline}
\end{table}

Although
communities have been studying
possible attack techniques against
heap vulnerabilities (see, \autoref{f:tbl-timeline}),
such techniques are often
considered arts, and thus
the approaches to discover them
remain ad-hoc, manual, and allocator-specific
at best.
Unfortunately,
such a trend
makes it hard for communities
to inherit or share lessons and efforts
in two dimensions,
namely, time and space.
%
In terms of time,
it is not easy for developers of heap allocators
to evaluate the feasibility of considered-to-be obsolete exploitation techniques
when introducing a new security or non-security feature.
For example,
a recent feature, called tcache in \ptmalloc,
that is designed to improve the performance of heap operations by introducing
a per-thread cache,
does not follow the common integrity checks of nearby chunks
during allocation or free,
rendering all existing security checks ineffective
against past exploitation techniques.
In terms of space,
it is difficult for developers of other heap allocators,
such as \dlmalloc, \jemalloc, and \tcmalloc,
to apply lessons from the communities of \ptmalloc
without spending a non-trivial amount of effort.
Not to mention,
it is not uncommon to implement a custom heap allocator
in systems software,
making it much harder to share such knowledge across them.

In this paper,
we present an automatic tool,
\sys,
to systematically discover
the unexplored heap exploitation primitives,
regardless of their underlying implementations.
The key idea of \sys
is to let the computer autonomously explore the spaces,
similar in concept to fuzzing,
which is proven to be practical and effective
in discovering software bugs~\cite{afl, syzkaller}.

However, it is non-trivial to apply classical fuzzing techniques
in discovering new heap exploitation primitives for three reasons.
First,
to successfully trigger a heap vulnerability,
it must generate a \emph{particular} sequence of steps
with exact data,
quickly rendering the problem intractable by using fuzzing approaches.
Accordingly,
researchers attempted
to tackle this problem
by using symbolic execution instead, 
but stumbled over the well-known state explosion problem,
thereby limiting its scope to
validating \emph{known} exploitation techniques~\cite{eckert:heaphopper}.
Second, 
we need to devise a fast way to estimate the possibility
of heap exploitation,
as fuzzing techniques require clear signals,
such as segmentation faults,
to recognize \emph{interesting} test cases.
Third,
the test cases generated by fuzzers
are typically redundant and obscure,
so users are required to spend
non-negligible time and effort
analyzing the final results.

The key intuition to overcome these challenges
(i.e., reducing search space)
is to abstract the internals of heap allocators
and the root causes of heap vulnerabilities
(see~\autoref{ss:generalization}).
In particular,
we observed that modern heap allocators
share three common design components,
namely,
\emph{binning},
\emph{in-place metadata},
and \emph{cardinal data}.
On top of these models,
we directed \sys to mutate and synthesize
heap operations and attack capabilities.
During the exploration,
\sys checks
whether the generated test case
can be potentially used
to construct exploitation primitives,
such as arbitrary write or overlapped chunks---%
we devised a notion called \emph{impacts of exploitation}
for efficient evaluation (see, \autoref{s:design:impact}).
Whenever \sys finds a new exploit primitive,
it generates as a proof a working PoC code
by using delta-debugging~\cite{zeller:delta-debugging}
to reduce the redundant test cases
to a minimal, equivalent class.


We evaluated \sys with three real-world allocators
(i.e., \ptmalloc, \tcmalloc, and \jemalloc)
as well as custom allocators
from the DARPA Cyber Grand Challenge.
As a result,
we discovered \emph{\newtech previously unknown}
exploitation techniques
against Linux's default heap allocator, \ptmalloc.
Compared with \heaphopper's approach,
which relies on symbolic execution to verify exploitation techniques,
\sys outperforms not just in
finding new techniques---none are found by \heaphopper ---
but also in validating known techniques
when no exploit-specific information is provided---%
only three out of eight techniques in \ptmalloc
were found by \heaphopper,
while \sys found them all.
While \heaphopper's approach
is limited to \ptmalloc (or its predecessor, \dlmalloc)
if no prior knowledge about a new allocator is available,
\sys's approach can be extended beyond \ptmalloc,
and indeed found exploit primitives
against other popular heap allocators,
such as \tcmalloc and \jemalloc,
as well as custom allocators from DARPA CGC.
To show the effectiveness of \sys's approach in other domains,
we also studied
how security features evolve
and which exploit primitives are effective
across different versions of \ptmalloc,
demonstrating the need
for an automated method
to evaluate the security of heap allocators.

In summary, we make the following contributions:
\squishlist
  \item
    We show that heap allocators share common designs,
    and define the impacts of exploitation,
    which can be used to efficiently evaluate exploitation techniques.
  \item We design, implement, and evaluate our prototype, \sys,
    the tool that automatically discovers heap exploitation techniques
    for various real-world allocators and custom allocators.
  \item
    \sys outperforms a state-of-the-art tool, \heaphopper, in
    finding new techniques and found \newtech new exploitation techniques in
    \ptmalloc and several techniques in \tcmalloc, \jemalloc, and custom
    allocators.
\squishend

%% file: fig/tbl-timeline.tex
\newcommand{\bulletpoint}{\makebox[0pt]{\textbullet}\hskip-0.5pt\vrule
width 1pt\hspace{\labelsep}}

\renewcommand{\arraystretch}{1.15}
\begin{tabular}{@{\,}r <{\hskip 2pt} !{\bulletpoint}
  >{\raggedright\arraybackslash} p{.72\columnwidth}}
  2001 & (1) Once upon a free()...~\cite{unlink}\\
  2003 & (1) Advanced Doug lea's malloc exploits~\cite{dl-exploit} \\
  2004 & (2) Exploiting the wilderness~\cite{wilderness} \\
  2007 & (2) The use of set_head to defeat the wilderness~\cite{set-head} \\
  2007 & (3) Understanding the heap by breaking it~\cite{ferguson:break-heap} \\
  2009 & (1) Yet another free() exploitation technique~\cite{yet-another-free} \\
  2009 & (6) Malloc Des-Maleficarum~\cite{malloc-des-maleficarum} \\
  2010 & (2) The house of lore: Reloaded~\cite{house-of-lore} \\
  2014 & (1) The poisoned {NUL} byte, 2014 edition \\
  2015 & (2) Glibc adventures: The forgotten chunk~\cite{the-forgotten-chunk} \\
  2016 & (3) Ptmalloc fanzine~\cite{ptmalloc-fanzine} \\
  2016 & (3) New exploit methods against Ptmalloc of Glibc~\cite{xie:new} \\
  2016 & (1) House of Einherjar~\cite{einherjar}\\
  \textbf{2018} & \textbf{(5) \sys}
\end{tabular}
\renewcommand{\arraystretch}{1}

%% file: background.tex
\section{Analysis of Heap Allocators}
\label{s:background}

\subsection{Modern Heap Allocators}
\label{s:background:allocators}

Dynamic memory allocation~\cite{lea:memory}
plays an essential role
in managing a program's heap space.
The C standard library
defines a set of APIs
to manage dynamic memory
allocations
such as \fmalloc and \ffree~\cite{kamp:malloc, malloc-man-page}.
For example, \fmalloc
allocates the given number of bytes
and returns a pointer to the allocated memory,
and \ffree reclaims
the memory specified by the given pointer.

A variety of heap allocators
have been developed
to meet the specific needs of target programs.
Heap allocators have two types of common goals:
\emph{good performance}
and \emph{small memory footprint}---%
minimizing the memory usage as well as
reducing fragmentation,
which is the unused memory (\ie hole) among in-use memory blocks.
Unfortunately,
these two desirable properties
are fundamentally conflicting;
an allocator should minimize
additional operations
to achieve good performance,
whereas it requires
additional operations
to minimize fragmentation.
Therefore,
the goal of an allocator
is typically to find
a good balance between these two goals
for its own workloads.

\PP{Common designs}
To achieve the aforementioned goals,
allocators share common designs:
\emph{binning},
\emph{in-place metadata},
and \emph{cardinal data}.

Many allocators
use size-based
classification,
known as binning.
They divide a whole size range into multiple groups
and manage memory blocks separately
according to their size group.
For example,
small-size memory blocks
focus on performance,
and
large-size memory blocks
focus on memory usage of the allocators.
Moreover,
by dividing size groups,
when they try to find the best-fit block
that is the smallest but sufficient block for given request,
they scan only blocks in the proper size group
instead of scanning all memory blocks.

Moreover,
many dynamic memory allocators
place metadata near the payload,
called \emph{in-place metadata}.
To minimize memory fragmentation,
a memory allocator
should maintain information
about allocated or freed memory
in metadata.
Even though
the allocator can
place metadata and payload
in distinct locations,
many allocators store
the metadata near the payload
to increase locality.
In particular,
by connecting metadata and payload,
an allocator can get benefits
from the cache.
Moreover,
in-place metadata
can reduce memory usage
by storing metadata
in the payload of freed memory.
Since the payload of freed
memory will no longer be used
by the application,
the allocator can reuse this part.

Further,
memory allocators
contain only
cardinal data
that are not encoded
and essential
for fast lookup and memory usage.
In particular,
metadata are
mostly
pointers or size-related values
that are used for
their data structures.
For example,
\ptmalloc
stores a raw pointer
for a linked list
that is used
to maintain freed memory blocks.

\begin{table}[!t]
 \centering
  \scriptsize
  \input{fig/tbl-other-allocators.tex}
  \caption{
    Common designs used in various memory allocators.
    This table shows that even though
    their detailed implementations could be different,
    heap allocators share common designs
    that can be exploited for automatic testing.
  }
  \label{f:tbl-other-allocators}
  \bm
\end{table}

\PP{Comparison of heap allocators}
%
To verify
whether memory allocators follow
common designs,
we manually investigated
widely used memory allocators,
\ptmalloc, \dlmalloc,
\jemalloc, \PartitionAlloc and \libumem,
as shown
in \autoref{f:tbl-other-allocators}.
All of the allocators use binning
and cardinal data
(\ie only pointers
and size-related information)
for their performance.
Many allocators still have used
in-place metadata
and some allocators have used
dedicated region for metadata
because of security concerns.
Their design decisions are various
based on their purpose,
\eg
\tcmalloc compromises security
for high performance,
whereas \PartitionAlloc~\cite{partitionalloc}
supports many security properties,
including isolation of objects
(\eg Bugs in DOM objects cannot corrupt JavaScript objects).
%

\begin{figure}[!t]
  \begin{subfigure}{\columnwidth}
    \input{code/ptmalloc-metadata.c}
    \coderule
  \end{subfigure}
  \begin{subfigure}{\columnwidth}
    \includegraphics[width=1\columnwidth]{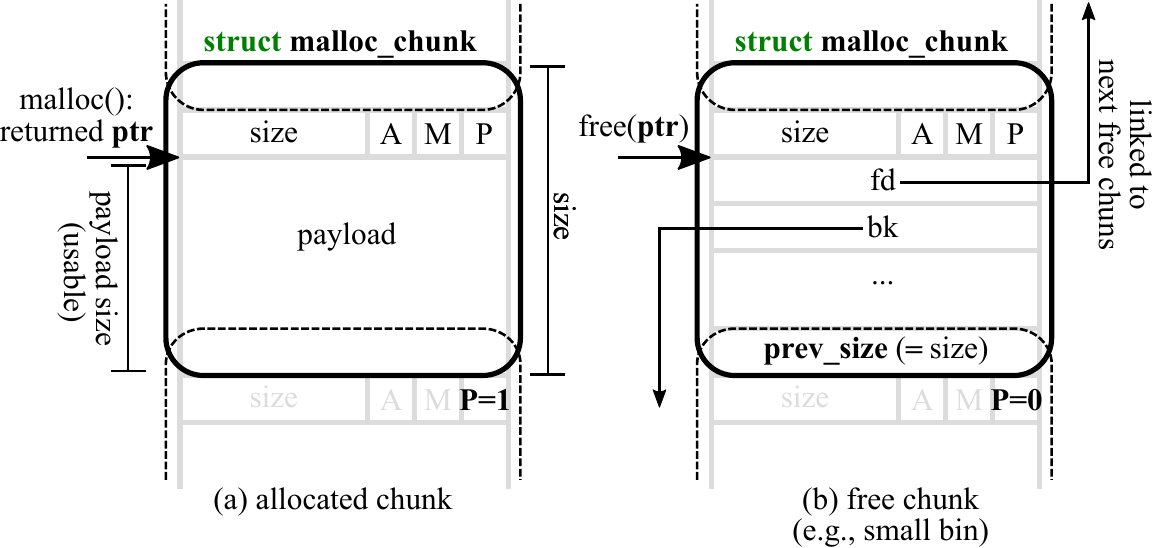}
  \end{subfigure}
  \caption{Metadata for a chunk in \ptmalloc
  and memory layout for the in-use and freed chunks~\cite{glibc-malloc-internals}.}
  \label{f:ptmalloc-metadata}
  \bm
\end{figure}

\subsection{ptmalloc: The Heap Allocator for glibc}
%
%
In this section,
we discuss \ptmalloc~\cite{glibc, ptmalloc, glibc-malloc-internals},
the heap allocator used in glibc,
whose exploitation techniques
have been heavily studied
due to its prevalence and its complexity of metadata%
~\cite{unlink, jemallocum, yet-another-free,
malloc-des-maleficarum, poison-null,
ferguson:break-heap, set-head, wilderness,
dl-exploit}.

\PP{Metadata}
A chunk in \ptmalloc
is a memory region containing metadata
and payload.
Memory allocation API such as \fmalloc
returns the address of the payload in the chunk.
\autoref{f:ptmalloc-metadata}
shows the metadata of a chunk
and its memory layout
for an in-use and a freed chunk.
\cc{prev_size} represents the size of a previous chunk
if it is freed.
We note that \cc{prev_size} of a chunk
is overlapped with the payload of the previous chunk.
This is legitimate since \cc{prev_size}
is considered only after the previous chunk is freed,
\ie the payload is no longer used.
\cc{size} represents the size of a current chunk.
The real size of the chunk is 8-bit aligned,
and the 3 LSBs of the size are used for
storing the state of the chunk.
The last bit of size,
called \previnuse (P),
shows whether
the previous chunk
is in-use.
For example,
in \autoref{f:ptmalloc-metadata},
after the chunk is freed,
the \previnuse in the next chunk
is changed from 1 to 0.
Other metadata,
\cc{fd}, \cc{bk}, \cc{fd_nextsize},
and \cc{bk_nextsize},
are used for maintaining linked lists
that hold freed chunks.

\PP{Binning}
\ptmalloc
has several types of bins:
fast bin, small bin, large bin,
unsorted bin, and tcache~\cite{tcache-benchmark},
which behaves like a caching layer
for allocation and free.
\autoref{f:tbl-ptmalloc-bin}
summarizes the characteristics of each bin.

\begin{table}[!t]
  \centering
  \footnotesize
  \input{fig/tbl-ptmalloc-bin.tex}

  \caption{
    The characteristics of bins
    in \ptmalloc in a 32/64-bit environment;
    the number of bins,
    range of size of bins,
    size consistency
    (\ie chunks in a bin has same size),
    what linked lists are maintained by the bin,
    and its merging.
    The sizes before the slash (/)
    in the range column
    are for a 32-bit environment,
    and the sizes after the slash
    are for a 64-bit environment.
  }
  \label{f:tbl-ptmalloc-bin}
  \bm
\end{table}

\ptmalloc has 10 fast bins
that store small freed chunks.
Since the chunks are not merged
and their sizes in the same fast bin are consistent,
\ptmalloc does not need to remove
a chunk of a fast bin in the middle.
Therefore, \ptmalloc uses
a single-linked list
that requires a smaller number of bookkeeping operations,
so it is faster than a double-linked list.
A fast bin maintains
the linked list
using \cc{fd} of the metadata.

Different from the fast bin,
a small bin allows merging
(aka., consolidation).
In \ffree,
a small-bin chunk
merges with other adjacent chunks
that are already freed.
This helps to reduce
memory fragmentation.
Due to merging,
a chunk in the middle of a small bin
needs to be removed
and added to another bin
for a larger size.
To support this chunk modification,
a small bin manages chunks
using a doubly-linked list
defined by \cc{fd} and \cc{bk} in the chunks.

A large bin is similar to the small bin
but can have variable-size chunks
in a single bin.
Therefore,
unlike a fast bin or a small bin
that can find the best-fit chunk
in a constant time
by accessing the first entry of the bin,
a large bin requires
scanning its list to find one.
To optimize this scan,
a large bin maintains
another sorted, double-linked list
that is defined by the metadata,
\cc{fd_nextsize} and \cc{bk_nextsize}.

The unsorted bin is a special bin
that serves as a fast, staging place for free chunks.
If a small or large chunk is freed,
it first moves to the unsorted bin.
When allocating memory,
\ptmalloc first scans the unsorted bin
to find a chunk
before scanning other bins.
During this scan, if a chunk in
the unsorted bin is not suitable
for allocation, it will move to a regular bin
(\ie a small bin or a large bin).
Using the unsorted bin,
\ptmalloc can defer
the decision for the regular bins
and increase locality
to improve performance.

The tcache, per-thread cache,
is enabled
by default from glibc 2.26.
It works similar to a fast bin
but requires no locking as allocated per thread,
and therefore it can achieve significant
performance improvements for multithread programs%
~\cite{tcache-benchmark}.

\PP{Consolidation}
Unlike a chunk in a non-fast bin,
a chunk in a fast bin
does not consolidate when freed.
Instead,
\ptmalloc
consolidates all freed fast bin chunks at once
using \consolidate in special cases,
for example,
allocating large bin size memory in \fmalloc%
~\cite{ptmalloc-fanzine}.
%
Since these cases are rare,
\ptmalloc can improve the performance of a fast bin
by deferring its consolidation
as much as possible.

\PP{Special chunks}
%
The top chunk (aka., the wilderness chunk)
is a special chunk
that borders the top of
the system memory.
Because of its location,
the top chunk is the only chunk
that is extendable
using the \cc{sbrk} system call.
To avoid fragmentation,
the top chunk is used
to serve a memory allocation request
only if no other chunk can serve
the request.

\subsection{Security Checks in ptmalloc}
\label{s:bg:security}
%
%
%


%
%
\begin{table}[t]
  \centering
  \footnotesize
  \input{fig/tbl-security-checks.tex}

  \caption{
    Security checks in \ptmalloc;
    a check's name that consists of
    its type and a unique identifier,
    an error message for its failure,
    and version that the check is first introduced,
    and covered checks by \sys in Ubuntu versions
    (details in \autoref{s:eval:abort}).
  }
  \label{f:tbl-security-checks}
  \bm
\end{table}
To prevent heap exploitation,
\ptmalloc
performs a lot of security checks,
verifying the integrity of heap metadata.
Whenever it finds a potential integrity violation of heap metadata,
it aborts the execution of a program
with an error message describing the detected violation.
To better understand these checks,
we categorize them into the following five groups,
as shown in \autoref{f:tbl-security-checks}.

\PP{Data structure integrity (D1--D6)}
The most dominant type of checks in \ptmalloc
is to verify the integrity of internal data structures.
In particular,
they check the structure of a double-linked list;
for example, a next link of one node's previous link
should point to the node itself.
Since this invariant should be satisfied
in the lifetime of all double-linked lists,
\ptmalloc performs this check in many places
whenever possible,
such as
\cc{unlink()} (D1, D2),
\ffree (D3),
and \fmalloc (D4, D5, D6).
Note that it is possible to check the integrity of
the whole double-linked list by iterating all nodes,
but due to the performance concern,
\ptmalloc checks a corresponding chunk
that it is about to perform any operation on.

\PP{Size range constraints (S1--S4)}
A size value in \ptmalloc's metadata
has universal constraints:
the size should be greater than the minimum size
to contain metadata,
and it should be smaller than
the system memory size.
These security checks verify whether
a size value satisfies these constraints.

\PP{Freeable memory checks (F1--F4)}
To reduce fragmentation,
\ptmalloc maintains
information
related to free chunks.
By using this information,
\ptmalloc can check whether
the memory to free
is valid.
For example,
F1 checks whether the memory is already freed
using \previnuse of its next chunk---%
due to the consolidation,
there are no two contagious free chunks in \ptmalloc.
This also can be checked
using the latest freed chunk (F2),
the top chunk (F3), and the boundary of heap (F4).

\PP{Uniform size check (U1--U2)}
Since chunks in a fast bin and a small bin
must have the same size,
U1 and U2 check this invariant
in \fmalloc and \consolidate, respectively.

\PP{Specialized checks (SP1--SP5)}
SP1 compares \cc{sbrk} syscall with the top chunk,
and SP2 checks consistency between a chunk size
and its corresponding \cc{prev_size}.
Moreover,
SP3 validates a freeing pointer,
and SP4 checks page-alignment in \cc{munmap()}.
SP5 is distinct from others
since it checks the consistency of a fast bin in
{a multi-threaded environment}.

%
%

\begin{table*}[!t]
  \centering
  \input{fig/tbl-known.tex}
  \caption{
    Modern heap exploitation techniques
    from recent work~\cite{eckert:heaphopper}
    including new ones found by \sys
    in \ptmalloc with abbreviations
    and brief descriptions.
    For brevity,
    we omitted tcache-related techniques.
    %
  }
  \label{f:tbl-known}
  \bm
  \vspace{-3px}
\end{table*}

\begin{figure}
  \begin{subfigure}{\columnwidth}
    \input{code/new-unlink.h}
    \coderule
    \caption{
      Security checks introduced since glibc 2.2.4 and 2.26.
      Two security checks first validate two invariants
      (see, comments above) before unlinking
      the victim chunk (i.e., \cc{P}).
    }
    \label{f:new-unlink}
  \end{subfigure}
  \begin{subfigure}{\columnwidth}
    \input{code/old-unlink-exploit.c}
    \coderule
    \caption{The unsafe unlink exploitation in glibc 2.3.3}
    \label{f:old-unlink-exploit}
  \end{subfigure}
  \begin{subfigure}{\columnwidth}
    \input{code/new-unlink-exploit.c}
    \coderule
    \caption{The unsafe unlink exploitation in glibc 2.27~\cite{how2heap}}
    \label{f:new-unlink-exploit}
  \end{subfigure}

  \caption{
    The unlink macros and corresponding exploits
    in glibc 2.3.3 and glibc 2.27.
    Compared to glibc 2.3.3,
    two security checks have been added in glibc 2.27.
    The first one hardens the off-by-one overflow
    and the second one hardens unlinking abuse.
    %
    Even though the security checks harden the attack,
    it is still avoidable.
  }
  \label{f:unlink}
  \vspace{-9px}
\end{figure}

\subsection{Heap Exploitation}

If an attack found a vulnerability
that corrupts heap metadata (\eg overflow)
or improperly uses heap APIs (\eg double free),
the next step is to develop the bug
to do a more useful exploit primitive
such as arbitrary write.
To do so,
attackers typically have to modify the heap metadata,
craft a fake chunk,
or call other heap APIs
according to the implementation of the target heap allocator.
Unfortunately,
this development is far from trivial
since it requires
in-depth understanding of an allocator
not just to abuse its metadata
but to avoid all relevant security checks.
Therefore,
researchers have studied and shared
{heap exploitation techniques}
that are reusable methods
to develop a vulnerability to a useful attack primitive%
~\cite{unlink, jemallocum, yet-another-free,
  malloc-des-maleficarum, poison-null,
  ferguson:break-heap, set-head, wilderness,
  dl-exploit, xie:new,
  poison-null, einherjar}.
\autoref{f:tbl-known} shows
modern heap exploitation techniques
collected from previous work~\cite{eckert:heaphopper}
and new ones that \sys found.

\PP{Example: Unsafe unlink}
One of the most famous heap exploitation technique is
the \emph{unsafe unlink attack},
which abuses the unlink mechanism of a double-linked list
in heap allocators,
as illustrated in~\autoref{f:new-unlink} and~\autoref{f:old-unlink-exploit}.
By modifying a forward pointer (\cc{P->fd})
into a properly encoded target location
and a backward pointer (\cc{P->bk})
into a desired value,
attackers can write
the value to the target location
(\cc{P->fd->bk} = \cc{P->bk}).
Due to the prevalence of
a double-linked list,
the same technique had been used
for many allocators,
including \dlmalloc, \ptmalloc,
and even the Windows heap allocator~\cite{unlink}.

To mitigate this attack,
allocators added new security checks
in \autoref{f:new-unlink},
which turn out to be insufficient to prevent the attack.
The check
verifies an invariant
of a double-linked list
that a backward pointer
of a forward pointer of a chunk
should point to the chunk
(\ie \cc{P->fd->bk == P})
and vice versa.
Therefore,
attackers cannot make the pointers
directly refer to arbitrary locations
as before
since the pointers
will not hold the invariant.
Even though the check prevents
the aforementioned attack,
attackers can avoid this check
by making a fake chunk
meet the condition,
as in~\autoref{f:new-unlink-exploit}.
Compared to the previous one,
the check makes the exploitation more complicated,
but still feasible.

\subsection{Generalizing Heap Exploitation}
\label{ss:generalization}

Heap exploitation can be generalized in three aspects:
1) types of bugs
(i.e., allowing an attacker to divert the program into unexpected states),
2) capabilities of attackers
(i.e., defining legitimate actions for an attacker to launch),
and 3) impact of exploitation
(i.e., describing what an attacker can achieve as a result).
This section elaborates on each of these aspects.

\PP{1) Types of bugs.}
There are four common types of heap-related bugs
that instantiate exploitation:

\squishlist
\item \textbf{Overflow (OF):}
  Writing beyond an object boundary.
\item \textbf{Write-after-free (WF):}
  Reusing a freed object.
\item \textbf{Arbitrary free (AF):}
  Freeing an arbitrary pointer.
\item \textbf{Double free (FF):}
  Freeing a previously reclaimed object.
\squishend

\noindent
Each of theses mistakes of a developer
allows attackers to divert
the program into unexpected states
in a certain way:
\textbf{overflow} allows modification
of the all metadata
(e.g., \cc{struct malloc_chunk} in \autoref{f:ptmalloc-metadata})
of any consequent chunks
(e.g., freed or allocated objects or even special chunks like top);
\textbf{write-after-free} allows
modification of the free metadata
(e.g., \cc{fd/bk} in \autoref{f:ptmalloc-metadata}),
which is similar in spirit to use-after-free;
\textbf{double free} allows violation
of the operational integrity of the internal heap metadata
(e.g., multiple reclaimed pointers linked in the heap structure);
and \textbf{arbitrary free} similarly
breaks the operational integrity of the heap management
but in a highly controlled manner---%
freeing an object with the crafted metadata
(e.g., \cc{size} in \autoref{f:ptmalloc-metadata}).
%
Since \textbf{overflow} enables
a variety of paths for exploitation,
we further characterize its types
based on common mistakes and errors
by developers.

\squishlist
\item \textbf{Off-by-one (O1):}
  Overwriting the last byte
  of the next consequent chunk
  (e.g., when making a mistake in size calculation,
  such as CVE-2016-5180~\cite{cve-2016-5180}).
  It overwrites \cc{P}-bit of \cc{size} in~\autoref{f:ptmalloc-metadata}.
\item \textbf{Off-by-one NULL (O1N):}
  Similar to the previous type,
  but overwriting the NULL byte
  (e.g., when using string related libraries such as \cc{sprintf}).
  It overwrites \cc{P=1} to \cc{P=0} in~\autoref{f:ptmalloc-metadata},
  tricking the allocated object to be freed.
\squishend

\noindent
It is worth noting that, unlike
a typical exploit scenario
that assumes arbitrary read and writes
in heap exploitation,
we exclude such a primitive for two reasons:
it is too specific to applications and execution contexts,
hardly meaningful for generalization, and
it is often too powerful for attackers
to launch easier attacks,
demotivating heap exploitation.
Therefore,
such powerful primitives are rather considered
one of the ultimate goals
of heap exploitation.

\PP{2) Capabilities of attackers}
To commonly describe heap exploitation techniques,
we clarify legitimate actions
that an attacker can launch.
First, an attacker can allocate
an object with an \emph{arbitrary size},
and free the object in an \emph{arbitrary order}.
This essentially means that
the attack can invoke
an arbitrary number of \malloc
with an arbitrary size parameter
and invoke \free (or not) in whichever order
the attacker wishes.

Second, an attacker can \emph{write arbitrary data}
on legitimate memory regions
(i.e., the \cc{payload}
in~\autoref{f:ptmalloc-metadata}
or global memory).
Although such legitimate behaviors in theory
depend largely on applications,
complex, real-world applications
typically exhibit such behaviors.
For example, in browsers,
attackers can arbitrarily invoke \malloc with an arbitrary size
by allocating \cc{ArrayBuffer},
and \free these objects by reclaiming them.
In addition, the attacker can write any data into
the allocated \cc{ArrayBuffer}.
However, it is worth noting
that it is always more favorable to attackers
if a heap exploit technique requires
fewer capabilities
than what is described here,
and in such cases,
we make a side note for better clarification.

\PP{3) Impact of exploitation}
The goal of each heap exploitation technique
is to develop common types of heap-related bugs
into more powerful exploit primitives
for a full-fledged attack.
For the systematization of a heap exploit,
we categorize its final impact
(i.e. achieved exploit primitives)
into four classes:

\squishlist
\item \textbf{Arbitrary-chunk (AC):}
  Hijacking the next \malloc to return
  an arbitrary pointer of choice.
\item \textbf{Overlapping-chunk (OC):}
  Hijacking the next \malloc to return
  a chunk inside a controllable (e.g., over-writable)
  chunk by an attacker.
\item \textbf{Arbitrary-write (AW):}
  Developing the heap-related bug
  into an arbitrary write (a write-where-what primitive).
\item \textbf{Restricted-write (RW):}
  Similar to arbitrary-write,
  but with various restrictions
  (e.g., non-controllable ``what''
  but a static pointer to a global heap structure like bins).
\squishend

Attackers might want to launch
a control-hijacking attack
by using these exploit primitives
combined with application-specific execution contexts.
For example, in the unsafe unlink case
(see,~\autoref{f:unlink}),
attackers can develop a word-byte overflow
to the arbitrary write
by repeatedly referring and writing
the object from a local variable (i.e., \cc{p1}).

%% file: fig/tbl-other-allocators.tex
\begin{tabular}{lcccl}
  \toprule

  \textbf{Allocators}
  & \textbf{B}
  & \textbf{I}
  & \textbf{C}
  & \multicolumn{1}{c}{\textbf{Description (applications)}}\\
\midrule
  \ptmalloc & \V & \V & \V & A default allocator in Linux. \\
  \dlmalloc & \V & \V & \V & An allocator that \ptmalloc is based on. \\
  \jemalloc &  \V & & \V & A default allocator in FreeBSD. \\
  \tcmalloc & \V & \V & \V & A high-performance allocator from Google. \\
  \PartitionAlloc &  \V & & \V & A default allocator in Chromium. \\
  \libumem &  \V & & \V & A default allocator in Solaris. \\
\bottomrule
\end{tabular}

\vspace{5pt}
\textbf{B}: Binning,
\textbf{I}: In-place metadata,
\textbf{C}: Cardinal data

%% file: code/ptmalloc-metadata.c.tex
\begin{Verbatim}[commandchars=\\\{\},codes={\catcode`\$=3\catcode`\^=7\catcode`\_=8}]
\PY{k}{struct} \PY{n}{malloc\PYZus{}chunk} \PY{p}{\PYZob{}}
  \PY{c+c1}{// size of \PYZdq{}previous\PYZdq{} chunk}
  \PY{c+c1}{//  (only valid when the previous chunk is freed, P=0)}
  \PY{k+kt}{size\PYZus{}t} \PY{n}{prev\PYZus{}size}\PY{p}{;}  
  
  \PY{c+c1}{// size in bytes (aligned by double words): lower bits}
  \PY{c+c1}{// indicate various states of the current/previous chunk}
  \PY{c+c1}{//   A: alloced in a non\PYZhy{}main arena}
  \PY{c+c1}{//   M: mmapped}
  \PY{c+c1}{//   P: \PYZdq{}previous\PYZdq{} in use (i.e., P=0 means freed)}
  \PY{k+kt}{size\PYZus{}t} \PY{n}{size}\PY{p}{;}

  \PY{c+c1}{// double links for free chunks in small/large bins}
  \PY{c+c1}{//  (only valid when this chunk is freed)}
  \PY{k}{struct} \PY{n}{malloc\PYZus{}chunk}\PY{o}{*} \PY{n}{fd}\PY{p}{;}
  \PY{k}{struct} \PY{n}{malloc\PYZus{}chunk}\PY{o}{*} \PY{n}{bk}\PY{p}{;}
  
  \PY{c+c1}{// double links for next larger/smaller size in largebins}
  \PY{c+c1}{//  (only valid when this chunk is freed)}
  \PY{k}{struct} \PY{n}{malloc\PYZus{}chunk}\PY{o}{*} \PY{n}{fd\PYZus{}nextsize}\PY{p}{;}
  \PY{k}{struct} \PY{n}{malloc\PYZus{}chunk}\PY{o}{*} \PY{n}{bk\PYZus{}nextsize}\PY{p}{;}
\PY{p}{\PYZcb{}}\PY{p}{;}
\end{Verbatim}

%% file: fig/tbl-ptmalloc-bin.tex
\begin{tabular}{cccccc}
  \toprule

  \textbf{Name}
  & \textbf{Num}
  & \textbf{Range}
  & \textbf{Uniform}
  & \textbf{List}
  & \textbf{Merge} \\

\midrule
  TCache
  & 64
  & $[0, 516 / 1032]$
  & \V
  & 1 Single
  &  \\

\midrule
  Fast
  & 10
  & $[0, 64 / 128)$
  & \V
  & 1 Single
  &  \\

  Small
  & 62
  & $[0, 508 / 1016)$
  & \V
  & 1 Double
  & \V \\

  Large
  & 63
  & $[508 / 1016, \infty)$
  & 
  & 2 Double
  & \V \\

  Unsorted
  & 1
  & $[0, \infty)$
  &
  & 1 Double
  & \V \\

  \bottomrule
\end{tabular}

%% file: fig/tbl-security-checks.tex
\addtolength{\tabcolsep}{-3.3pt}

\resizebox{\columnwidth}{!}{\begin{tabular}{clccc}
\toprule
    \multicolumn{1}{c}{\textbf{Name}}
  & \multicolumn{1}{c}{\textbf{Error message}}
  & \multicolumn{1}{c}{\textbf{Version}}
  & \textbf{Xenial}
  & \textbf{Bionic} \\
\midrule
D1 & corrupted double-linked list                     &  2.3.4  & \V & \V \\
D2 & corrupted double-linked list (not small)         &  2.21   &    & \V \\
D3 & free(): corrupted unsorted chunks                &  2.11   & \V & \V \\
D4 & malloc(): corrupted unsorted chunks 1            &  2.11   &         \\
D5 & malloc(): corrupted unsorted chunks 2            &  2.11   & \V & \V \\
D6 & malloc(): smallbin double linked list corrupted  &  2.11   & \V & \V \\
S1 & free(): invalid next size (fast)                 &  2.3.4  & \V & \V \\
S2 & free(): invalid next size (normal)               &  2.3.4  & \V & \V \\
S3 & free(): invalid size                             &  2.4    & \V & \V \\
S4 & malloc(): memory corruption                      &  2.3.4  & \V & \V \\
F1 & double free or corruption (!prev)                &  2.3.4  & \V & \V \\
F2 & double free or corruption (fasttop)              &  2.3.4  & \V & \V \\
F3 & double free or corruption (top)                  &  2.3.4  & \V & \V \\
F4 & double free or corruption (out)                  &  2.3.4  & \V & \V \\
U1 & malloc(): memory corruption (fast)               &  2.3.4  & \V & \V \\
U2 & malloc_consolidate(): invalid chunk size         &  2.27   & --- & \V \\
SP1 & break adjusted to free malloc space             &  2.10.1 & \V & \V \\
SP2 & corrupted size vs. prev_size                    &  2.26   & \V & \V \\
SP3 & free(): invalid pointer                         &  2.0.1  & \V & \V \\
SP4 & munmap_chunk(): invalid pointer                 &  2.4    & \V & \V \\
SP5 & invalid fastbin entry (free)                    &  2.12.1 &    &    \\
\bottomrule
\end{tabular}}

\vspace{5pt}
\textbf{D}: Data structure integrity checks,
\textbf{S}: Size range constraints, \\
\textbf{F}: Freeable memory checks,
\textbf{U}: Uniform size checks,
\textbf{SP}: Specialized checks

\addtolength{\tabcolsep}{3.3pt}

%% file: fig/tbl-known.tex
\newcommand{\doublefree}{DF}
\newcommand{\overflow}{OV}
\newcommand{\writefree}{WF}
\newcommand{\arbitraryfree}{AF}
\newcommand{\offbyonenull}{ON}
\newcommand{\offbyone}{O1}

\newcommand{\overlapinheap}{OC}
\newcommand{\arbitraryloc}{AC}
\newcommand{\arbitrarywrite}{AW}
\newcommand{\restrictedwrite}{RW}

\resizebox{\textwidth}{!}{\begin{tabular}{llp{345px}l}
  \toprule
  \textbf{Name}
  & \textbf{Abbr.} & \textbf{Description}
  & \textbf{New} \\
  \midrule
    Fast bin dup & FD
                 & Corrupting a fast bin freelist (\eg by double free or write-after-free)
                    to return an arbitrary location \\
    Unsafe unlink & UU
      & Abusing unlinking in a freelist to get arbitrary write \\
    House of spirit & HS
      & Freeing a fake chunk of fast bin to return arbitrary location \\
    Poison null byte & PN
      & Corrupting heap chunk size to consolidate chunks even in the presence of
      allocated heap \\
    House of lore & HL
      & Abusing the small bin freelist to return an arbitrary location \\
     Overlapping chunks & OC
       & Corrupting a chunk size in the unsorted bin to overlap with an
       allocated heap \\
    House of force & HF
      & Corrupting the top chunk to return an arbitrary location \\
    Unsorted bin attack & UB
      & Corrupting a freed chunk in unsorted bin to write a uncontrollable
      value to arbitrary location \\
    House of einherjar & HE
      & Corrupting \previnuse to consolidate chunks
      to return an arbitrary location that requires a heap address \\
    \midrule
      Unsorted bin into stack & UBS
      & Abusing the unsorted freelist to return an arbitrary location & \V \\

      House of unsorted einherjar & HUE
      & A variant of house of einherjar that does not require a heap address & \V \\
      Unaligned double free & UFF
      & Corrupting a small bin freelist to return already allocated heap & \V \\
      Overlapping small chunks & OCS
      & Corrupting a chunk size in a small bin to overlap chunks & \V \\
      Fast bin into other bin & FDO
      & Corrupting a fast bin freelist and use \consolidate
      to return an arbitrary non-fast-bin chunk & \V \\
  \bottomrule
\end{tabular}}

%% file: code/new-unlink.h.tex
\begin{Verbatim}[commandchars=\\\{\},codes={\catcode`\$=3\catcode`\^=7\catcode`\_=8}]
\PY{k+kt}{\PYZsh{}define} \PY{n+nf}{unlink}\PY{p}{(}\PY{n}{AV}\PY{p}{,} \PY{n}{P}\PY{p}{,} \PY{n}{BK}\PY{p}{,} \PY{n}{FD}\PY{p}{)}                                    \PYZbs{}
    \PY{c+cm}{/* (1) checking if size == the next chunk\PYZsq{}s prev\PYZus{}size */}     \PYZbs{}
$\star$   \PY{k}{if} \PY{p}{(}\PY{n}{chunksize}\PY{p}{(}\PY{n}{P}\PY{p}{)} \PY{o}{!}\PY{o}{=} \PY{n}{prev\PYZus{}size}\PY{p}{(}\PY{n}{next\PYZus{}chunk}\PY{p}{(}\PY{n}{P}\PY{p}{)}\PY{p}{)}\PY{p}{)}                \PYZbs{}
$\star$     \PY{n}{malloc\PYZus{}printerr}\PY{p}{(}\PY{l+s}{\PYZdq{}}\PY{l+s}{corrupted size vs. prev\PYZus{}size}\PY{l+s}{\PYZdq{}}\PY{p}{)}\PY{p}{;}           \PYZbs{}
    \PY{n}{FD} \PY{o}{=} \PY{n}{P}\PY{o}{\PYZhy{}}\PY{o}{\PYZgt{}}\PY{n}{fd}\PY{p}{;}                                                  \PYZbs{}
    \PY{n}{BK} \PY{o}{=} \PY{n}{P}\PY{o}{\PYZhy{}}\PY{o}{\PYZgt{}}\PY{n}{bk}\PY{p}{;}                                                  \PYZbs{}
    \PY{c+cm}{/* (2) checking if prev/next chunks correctly point to me */} \PYZbs{}
$\star$   \PY{n+nf}{if} \PY{p}{(}\PY{n}{FD}\PY{o}{\PYZhy{}}\PY{o}{\PYZgt{}}\PY{n}{bk} \PY{o}{!}\PY{o}{=} \PY{n}{P} \PY{o}{|}\PY{o}{|} \PY{n}{BK}\PY{o}{\PYZhy{}}\PY{o}{\PYZgt{}}\PY{n}{fd} \PY{o}{!}\PY{o}{=} \PY{n}{P}\PY{p}{)}                              \PYZbs{}
$\star$     \PY{n}{malloc\PYZus{}printerr}\PY{p}{(}\PY{l+s}{\PYZdq{}}\PY{l+s}{corrupted double\PYZhy{}linked list}\PY{l+s}{\PYZdq{}}\PY{p}{)}\PY{p}{;}           \PYZbs{}
$\star$   \PY{k}{else} \PY{p}{\PYZob{}}                                                       \PYZbs{}
      \PY{n}{FD}\PY{o}{\PYZhy{}}\PY{o}{\PYZgt{}}\PY{n}{bk} \PY{o}{=} \PY{n}{BK}\PY{p}{;}                                               \PYZbs{}
      \PY{n}{BK}\PY{o}{\PYZhy{}}\PY{o}{\PYZgt{}}\PY{n}{fd} \PY{o}{=} \PY{n}{FD}\PY{p}{;}                                               \PYZbs{}
      \PY{p}{.}\PY{p}{.}\PY{p}{.}                                                        \PYZbs{}
$\star$    \PY{p}{\PYZcb{}}
\end{Verbatim}

%% file: code/old-unlink-exploit.c.tex
\begin{Verbatim}[commandchars=\\\{\},codes={\catcode`\$=3\catcode`\^=7\catcode`\_=8}]
  \PY{c+c1}{// [PRE\PYZhy{}CONDITION]}
  \PY{c+c1}{//   sz : any non\PYZhy{}fast\PYZhy{}bin size}
  \PY{c+c1}{//   dst: where to write (void*)}
  \PY{c+c1}{//   val: target value (ptr to writable memory)}
  \PY{c+c1}{// [BUG] buffer overflow (p1)}
  \PY{c+c1}{// [POST\PYZhy{}CONDITION]}
  \PY{c+c1}{//   *dst = val}
  \PY{k+kt}{void} \PY{o}{*}\PY{n}{p1} \PY{o}{=} \PY{n}{malloc}\PY{p}{(}\PY{n}{sz}\PY{p}{)}\PY{p}{;}
  \PY{k+kt}{void} \PY{o}{*}\PY{n}{p2} \PY{o}{=} \PY{n}{malloc}\PY{p}{(}\PY{n}{sz}\PY{p}{)}\PY{p}{;}

  \PY{k}{struct} \PY{n}{malloc\PYZus{}chunk} \PY{o}{*}\PY{n}{c2} \PY{o}{=} \PY{n}{raw\PYZus{}to\PYZus{}chunk}\PY{p}{(}\PY{n}{p2}\PY{p}{)}\PY{p}{;}

  \PY{c+c1}{// [BUG] overflowing p1}
  \PY{n}{c2}\PY{o}{\PYZhy{}}\PY{o}{\PYZgt{}}\PY{n}{prev\PYZus{}size} \PY{o}{=} \PY{l+m+mi}{0}\PY{p}{;}
  \PY{c+c1}{// next chunk\PYZsq{}s size == c2\PYZsq{}s prev\PYZus{}size, tricking c2 freed (P=0)}
  \PY{n}{c2}\PY{o}{\PYZhy{}}\PY{o}{\PYZgt{}}\PY{n}{size} \PY{o}{=} \PY{o}{\PYZhy{}}\PY{k}{sizeof}\PY{p}{(}\PY{k+kt}{void}\PY{o}{*}\PY{p}{)}\PY{p}{;}
  \PY{n}{c2}\PY{o}{\PYZhy{}}\PY{o}{\PYZgt{}}\PY{n}{fd} \PY{o}{=} \PY{n}{dst} \PY{o}{\PYZhy{}} \PY{n}{offsetof}\PY{p}{(}\PY{k}{struct} \PY{n}{malloc\PYZus{}chunk}\PY{p}{,} \PY{n}{bk}\PY{p}{)}\PY{p}{;}
  \PY{n}{c2}\PY{o}{\PYZhy{}}\PY{o}{\PYZgt{}}\PY{n}{bk} \PY{o}{=} \PY{n}{val}\PY{p}{;}

  \PY{c+c1}{// trigger unlink(c2) via forward consolidation}
  \PY{n}{free}\PY{p}{(}\PY{n}{p1}\PY{p}{)}\PY{p}{;}

  \PY{n}{assert}\PY{p}{(}\PY{o}{*}\PY{n}{dst} \PY{o}{=}\PY{o}{=} \PY{n}{val}\PY{p}{)}\PY{p}{;}
\end{Verbatim}

%% file: code/new-unlink-exploit.c.tex
\begin{Verbatim}[commandchars=\\\{\},codes={\catcode`\$=3\catcode`\^=7\catcode`\_=8}]
  \PY{c+c1}{// Same PRE/POST CONDITIONS and BUG as (b)}
  \PY{k+kt}{void} \PY{o}{*}\PY{n}{p1} \PY{o}{=} \PY{n}{malloc}\PY{p}{(}\PY{n}{sz}\PY{p}{)}\PY{p}{;}
  \PY{k+kt}{void} \PY{o}{*}\PY{n}{p2} \PY{o}{=} \PY{n}{malloc}\PY{p}{(}\PY{n}{sz}\PY{p}{)}\PY{p}{;}

  \PY{k}{struct} \PY{n}{malloc\PYZus{}chunk} \PY{o}{*}\PY{n}{fake} \PY{o}{=} \PY{n}{p1}\PY{p}{;}
  \PY{c+c1}{// bypassing (1): P\PYZhy{}\PYZgt{}size == next\PYZus{}chunk(P)\PYZhy{}\PYZgt{}prev\PYZus{}size}
  \PY{n}{fake}\PY{o}{\PYZhy{}}\PY{o}{\PYZgt{}}\PY{n}{size} \PY{o}{=} \PY{k}{sizeof}\PY{p}{(}\PY{k+kt}{void}\PY{o}{*}\PY{p}{)}\PY{p}{;}
  \PY{c+c1}{// bypassing (2): P\PYZhy{}\PYZgt{}fd\PYZhy{}\PYZgt{}bk == P \PYZam{}\PYZam{} P\PYZhy{}\PYZgt{}bk\PYZhy{}\PYZgt{}fd == P}
  \PY{n}{fake}\PY{o}{\PYZhy{}}\PY{o}{\PYZgt{}}\PY{n}{fd} \PY{o}{=} \PY{p}{(}\PY{k+kt}{void}\PY{o}{*}\PY{p}{)}\PY{o}{\PYZam{}}\PY{n}{p1} \PY{o}{\PYZhy{}} \PY{n}{offsetof}\PY{p}{(}\PY{k}{struct} \PY{n}{malloc\PYZus{}chunk}\PY{p}{,} \PY{n}{bk}\PY{p}{)}\PY{p}{;}
  \PY{n}{fake}\PY{o}{\PYZhy{}}\PY{o}{\PYZgt{}}\PY{n}{bk} \PY{o}{=} \PY{p}{(}\PY{k+kt}{void}\PY{o}{*}\PY{p}{)}\PY{o}{\PYZam{}}\PY{n}{p1} \PY{o}{\PYZhy{}} \PY{n}{offsetof}\PY{p}{(}\PY{k}{struct} \PY{n}{malloc\PYZus{}chunk}\PY{p}{,} \PY{n}{fd}\PY{p}{)}\PY{p}{;}

  \PY{k}{struct} \PY{n}{malloc\PYZus{}chunk} \PY{o}{*}\PY{n}{c2} \PY{o}{=} \PY{n}{raw\PYZus{}to\PYZus{}chunk}\PY{p}{(}\PY{n}{p2}\PY{p}{)}\PY{p}{;}

  \PY{c+c1}{// [BUG] overflowing p1: it shrinks the previous chunk\PYZsq{}s size,}
  \PY{c+c1}{// tricking `fake\PYZsq{} as the previous chunk}
  \PY{n}{c2}\PY{o}{\PYZhy{}}\PY{o}{\PYZgt{}}\PY{n}{prev\PYZus{}size} \PY{o}{=} \PY{n}{chunk\PYZus{}size}\PY{p}{(}\PY{n}{sz}\PY{p}{)} \PYZbs{}
                  \PY{o}{\PYZhy{}} \PY{n}{offsetof}\PY{p}{(}\PY{k}{struct} \PY{n}{malloc\PYZus{}chunk}\PY{p}{,} \PY{n}{fd}\PY{p}{)}\PY{p}{;}
  \PY{c+c1}{// tricking the previous chunk freed, P=0}
  \PY{n}{c2}\PY{o}{\PYZhy{}}\PY{o}{\PYZgt{}}\PY{n}{size} \PY{o}{\PYZam{}}\PY{o}{=} \PY{o}{\PYZti{}}\PY{l+m+mi}{1}\PY{p}{;}

  \PY{c+c1}{// triggering unlink(fake) via backward consolidation}
  \PY{n}{free}\PY{p}{(}\PY{n}{p2}\PY{p}{)}\PY{p}{;}

  \PY{n}{assert}\PY{p}{(}\PY{n}{p1} \PY{o}{=}\PY{o}{=} \PY{p}{(}\PY{k+kt}{void}\PY{o}{*}\PY{p}{)}\PY{o}{\PYZam{}}\PY{n}{p1} \PY{o}{\PYZhy{}} \PY{n}{offsetof}\PY{p}{(}\PY{k}{struct} \PY{n}{malloc\PYZus{}chunk}\PY{p}{,} \PY{n}{bk}\PY{p}{)}\PY{p}{)}\PY{p}{;}
  \PY{c+c1}{// writing with p1: overwriting itself to dst}
  \PY{o}{*}\PY{p}{(}\PY{k+kt}{void}\PY{o}{*}\PY{o}{*}\PY{p}{)}\PY{p}{(}\PY{n}{p1} \PY{o}{+} \PY{n}{offsetof}\PY{p}{(}\PY{k}{struct} \PY{n}{malloc\PYZus{}chunk}\PY{p}{,} \PY{n}{bk}\PY{p}{)}\PY{p}{)} \PY{o}{=} \PY{n}{target}\PY{p}{;}
  \PY{c+c1}{// writing with p1: overwriting *dst with val}
  \PY{o}{*}\PY{p}{(}\PY{k+kt}{void}\PY{o}{*}\PY{o}{*}\PY{p}{)}\PY{n}{p1} \PY{o}{=} \PY{p}{(}\PY{k+kt}{void}\PY{o}{*}\PY{p}{)}\PY{n}{val}\PY{p}{;}

  \PY{n}{assert}\PY{p}{(}\PY{o}{*}\PY{n}{dst} \PY{o}{=}\PY{o}{=} \PY{n}{val}\PY{p}{)}\PY{p}{;}
\end{Verbatim}

%% file: challenge.tex
\section{Technical Challenges}
\label{s:challenge}

Our goal is to automatically explore
new types of heap exploitation techniques
given an implementation of any heap allocators---%
its source code is not required.
Such a capability not only enables
automatic exploit synthesis
but also makes several, unprecedented applications possible:
1) systematically discovering unknown types
 of heap exploitation schemes;
2) comprehensively evaluating
 the security of popular heap allocators;
and 3) providing insight into what and how to improve their security.
However, achieving this autonomous capability
is far from trivial, for the following reasons.

\PP{Autonomous reasoning of the heap space}
To find heap exploitation techniques,
we should handle a large search space
consisting of
enormous possible orders, arguments for heap APIs,
and data in the heap and global buffer.
This space could be greatly reduced
using exploit-specific knowledge~\cite{eckert:heaphopper},
however,
this is not applicable for finding new exploit techniques.
To resolve this issue,
we use a \emph{random} search algorithm
that is effective in exploring a large search space~\cite{heelan:automatic}.
In particular,
we use a fuzzer as a meta-explorer
by encoding heap actions from a binary form
that a fuzzer can mutate and synthesize effectively.
We also abstract common designs
of modern heap allocators
to further reduce the search space
(\autoref{s:design:actions}).

%
%


\PP{Devising exploitation techniques}
While enumerating possible candidates
for an exploit technique,
a system needs to verify
whether the candidate is valuable.
One way
to assess the candidates
is to synthesize
a full exploit automatically
(\eg spawning a shell),
but it is extremely difficult and inefficient,
especially for heap vulnerabilities%
~\cite{avgerinos:aeg, cha:mayhem, schwartz:q, heelan:automatic, repel:modular, chriseagle:cgc}.
To resolve this issue,
we devise the concept of
\emph{impact of exploitation}.
In particular,
we estimate the impacts of heap exploitation primitives
(i.e., AC, OC, AW, and RW)
during exploration instead of synthesizing a full exploit.
We will show that
these impacts
can be quickly detectable at runtime
by utilizing \emph{shadow memory}
(\autoref{s:design:impact}).
%
%

\PP{Normalization}
%
Even though
random search
is effective in exploring
a large search space,
an exploitation technique
found by this algorithm
tends to be redundant and inessential,
requiring non-trivial time
to analyze the result.
To fix this issue,
we leverage
\emph{delta-debugging} techniques to
minimize the redundant actions
and transform the found result
into an essential class.
This is so effective that
we could reduce \reducedactions of actions,
drastically helping us
to share the new exploitation techniques
with the communities
(\autoref{s:design:delta}).

%% file: design.tex
\section{Autonomous Exploration of Heap Exploitation}
\label{s:design}

\begin{figure}[!t]
  \centering
  \includegraphics[width=\columnwidth]{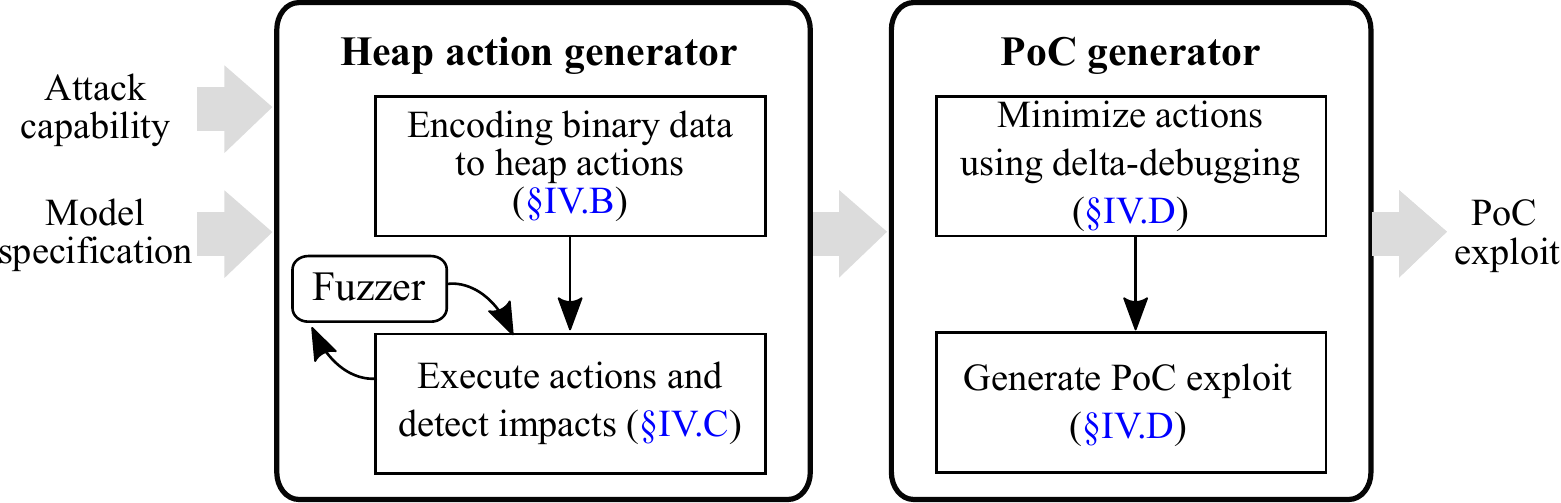}
  \caption{
    Overview of \sys.
    It first generates a sequence of heap actions
    given a model specification.
    While executing the generated actions,
    it estimates the impact of exploitation.
    Whenever a new exploit is found,
    it minimizes the actions and produces a PoC code.
  }
  \label{f:arch}
  \bm
\end{figure}

%
%
%
%
%
%
%

\subsection{Overview}

\begin{figure*}[!t]
  \begin{minipage}{.7\columnwidth}
    \includegraphics[width=\columnwidth]{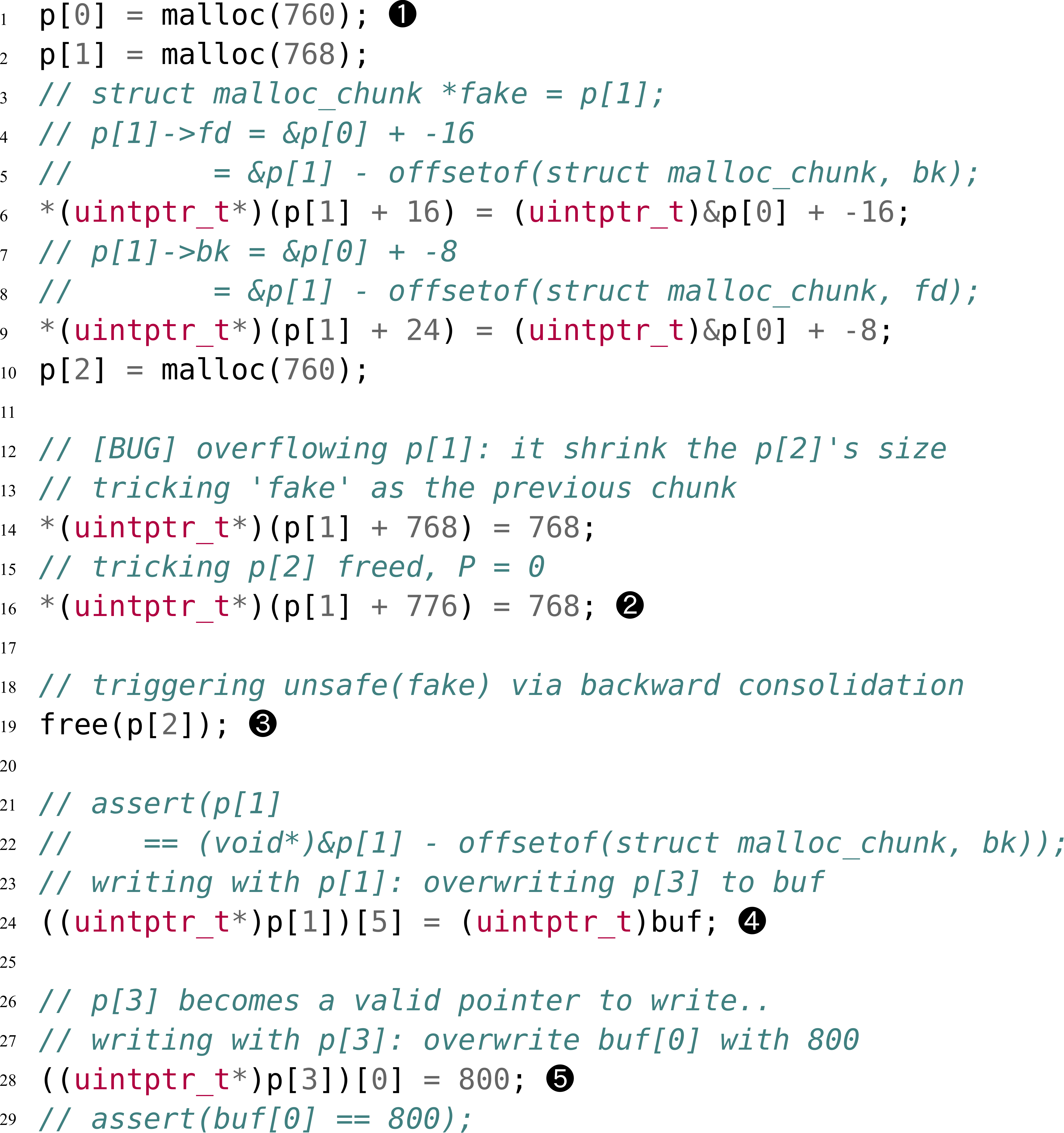}
    \caption{
      A PoC code of unsafe unlink found by \sys
      that has been simplified for easier explanation.
    }
    \label{f:unsafe-unlink-poc}
  \end{minipage}
  \hspace{3px}
  \vline
  \hspace{3px}
  \begin{minipage}{1.25\columnwidth}
    \includegraphics[width=\columnwidth]{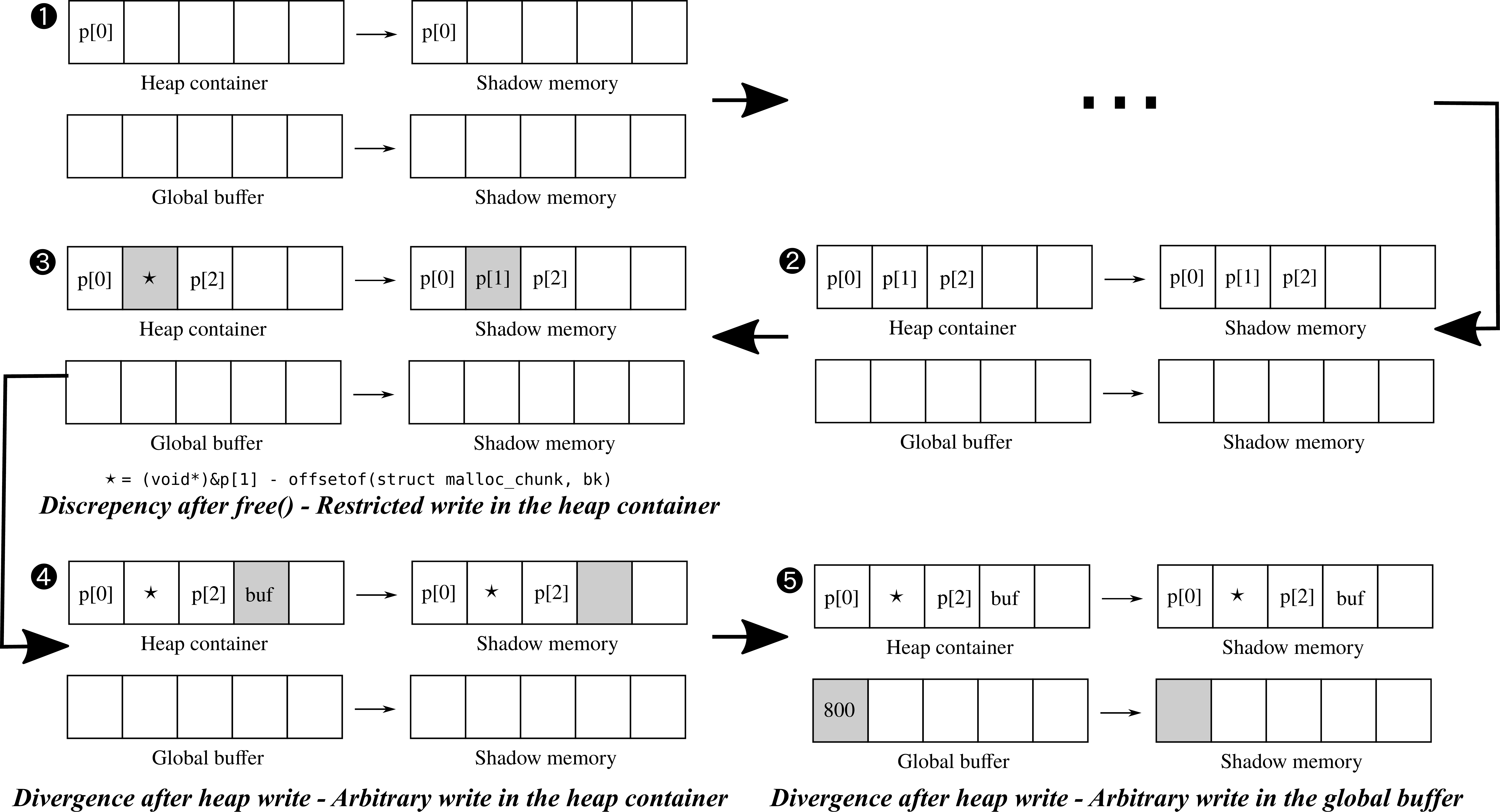}
    \caption{
      Shadow memory states in \autoref{f:unsafe-unlink-poc}.
      Black circles in left top corner
      represent locations in the code of states.
      Gray-color boxes show divergence between original memory and its
      shadow memory.
      Using this information,
      \sys can detect exploitation techniques.
    }
    \label{f:shadow-memory}
  \end{minipage}
  \bm
\end{figure*}

\sys follows a common paradigm
in classical fuzzing---%
test generation, crash detection, and test reduction,
but tailored to heap exploitation
(see \autoref{f:arch}).
It first mutates and generates
a sequence of heap actions
based on a user-provided model specification.
%
%
Heap actions that \sys can formulate
include heap allocation, free,
buffer writes, heap writes,
and bug invocation
(\autoref{s:design:actions}).
During execution,
\sys evaluates
whether the executed test case
results in impacts of exploitation,
similar in concept to detecting a crash in fuzzing
(\autoref{s:design:impact}).
Whenever \sys finds a new exploit,
it minimizes the heap actions
and produces as a proof
a PoC code (see, \autoref{f:unsafe-unlink-poc})
that contains only an essential set of actions
(\autoref{s:design:delta}).

\PP{Model specifications}
Users can optionally provide
a model specification
either to direct \sys to focus on a certain type of exploitation techniques
or to restrict the conditions
of a target environment.
For example,
\emph{house-of-force} (see~\autoref{f:tbl-known})
requires \emph{arbitrary} size allocation,
so without guidance,
\sys tends to converge
to another exploitation technique.
It accepts five types of restrictions:
chunk sizes, bugs, impacts, actions, and knowledge.
The first four types are self-explanatory,
and the knowledge is about
the ability of an attacker to break ASLR
(\ie prior knowledge of certain addresses).
Users can specify three types of addresses
that an attacker may know:
a global buffer address,
a heap address,
and a container address.

%
%

\subsection{Generating Actions for Abstract Heap Models}
\label{s:design:actions}

\sys generates five types of heap-related actions:
allocation, deallocation,
buffer writes,
heap writes,
and bug invocation.
It encodes each action
as a sequence of bits
such that the random output of a fuzzer
can be appropriately mapped
to these actions for interpretation.
To reduce the search space,
it formulates each action
on top of an \emph{abstract} heap model
that accommodates the common design idioms
of modern heap allocators.
The following explains
how each action
takes advantage
of the abstract model
in reducing the search space.


\PP{Allocation}
\sys selects the size of heap objects
based on the boundary values of
each \emph{bins} (see \autoref{s:background:allocators}).
Since
a heap allocator
manages each bin individually,
\sys needs different kinds of objects
to examine their different logic.
Thus,
\sys allocates memory
in random size,
but considering binning
(\textbf{I3}).
In particular,
\sys first randomly selects
a group of size
and then allocates an object
whose size is in this group.
The group is separated by
approximate boundary values
instead of implementation-specific ones
to make \sys compatible to
any allocator.
Currently,
\sys uses five boundaries
with exponential distance
from $2^0$ to $2^{20}$.
This division is arbitrary,
but sufficient for increasing the chances
to explore various bins.
%
%
Moreover,
\sys attempts to allocate
multiple objects in the same bin
(\textbf{I4}, \textbf{I5})
since an object interacts with other objects
only in the same bin.
For example,
in \ptmalloc,
a non-fast-bin object
merges with a non-fast-bin object,
not with a fast bin object.
To cover this interaction,
\sys allocates an object
whose size is related to
other objects' sizes.

To find certain techniques,
\sys also needs specialized sizes
(\textbf{I1}, \textbf{I2}).
For example,
a difference between an object and a buffer address
is required to allocate memory in the buffer
if integer overflow exists in an allocator.
Thus,
\sys also uses
several pre-defined constants
and differences between pointers
as its allocation size.

After selecting a size of an object,
\sys claims the object using \fmalloc API
and stores the object's address, size, and status (\ie allocated)
into its internal data structure, called \emph{the heap container}.
This information about an object
is used to perform other actions,
\eg deallocation or bug invocation.
%

%
%

\PP{Deallocation}
\sys deallocates
a randomly selected heap pointer
from the heap container.
To ensure
that this does not trigger a double free bug,
which will be emulated in the subsequent bug invocation action,
\sys checks an object's status.
If \sys chooses an already freed pointer,
it simply ignores the deallocation action
to avoid the bug.

\PP{Heap \& Buffer write}
%
To overcome limitations of fuzzing,
\sys profits from common designs
of heap allocators.
To find an exploit technique,
\sys needs to write
accurate data in order
to either heap or a controllable region
(\ie \emph{the global buffer} in \sys),
but such a task
is difficult for classical fuzzing.
Thus,
\sys exploits the in-place
and cardinal data of allocators
(see, \autoref{s:background:allocators})
to prune its search space
by limiting locations and data range, respectively.
In more detail,
\sys writes
only eight-word values
from the start or the end of an object
since a heap allocator
stores its metadata
near boundary for locality (in-place metadata).
Further,
\sys generates random values
(see, \autoref{f:tbl-strategy})
that can be used for sizes or pointers
in an allocator
instead of fully random ones (cardinal data).

To explore various
exploit techniques,
\sys introduces systematic noises
to generated values.
In particular,
\sys modifies a value
using linear (addition and multiplication) or
shift transformation (addition only)
according to the type of a value.
For example,
linear transformation
is prohibited to a pointer type
since multiplying a constant to a pointer is meaningless.
Moreover,
\sys considers
the alignment of pointer-type values
to further reduce its search space.
Similar to deallocation,
\sys writes data only
in a valid heap region
(\ie neither overflow nor underflow)
to ensure legitimacy of this action.

%
\begin{table}[t]
  \centering
  \footnotesize
  \input{fig/tbl-strategy.tex}
  \caption{
    Random values generated by \sys.
    \sys has two types of values:
    the integer type and the pointer type.
    \sys also defines
    an alignment requirement
    and transformation
    according to characteristics of each value.
  }
  \label{f:tbl-strategy}
  \bm
\end{table}


\PP{Bug invocation}
To explore exploitation techniques,
\sys needs to conduct buggy actions.
Currently,
\sys handles six bugs that are related to heap:
\textcircled{1} overflow,
\textcircled{2} write-after-free,
\textcircled{3} off-by-one overflow,
\textcircled{4} off-by-one NULL overflow,
\textcircled{5} double free,
and \textcircled{6} arbitrary free.
%

\sys performs only one of these bugs
for one technique
to limit the power of an adversary.
%
Therefore,
if a bug invocation action is provided
that is different from a previously executed one,
\sys simply ignores it.
However, \sys allows repetitive execution of the same bug
to emulate the situation in which
an attacker re-triggers the bug.

\sys deliberately builds a buggy action
to ensure its occurrence.
For overflow and off-by-one-bugs,
\sys uses the \cc{malloc_usable_size} API
to get the actual heap size
to calculate required size for overflow.
This is necessary
since the request size
could be smaller than the actual size
due to alignment or the minimum size constraint.
Specially for \ptmalloc,
\sys uses a dedicated single-line routine
to get the actual chunk size
since \ptmalloc's \cc{malloc_usable_size()}
is inaccurate under the presence
of memory corruption bugs.
Moreover, in the double free
and the write-after-free bugs,
\sys checks whether
a target chunk is already freed.
If it is not freed yet,
\sys ignores this buggy action
and waits for the next one.

\subsection{Detecting Exploitation Techniques by Impact}
\label{s:design:impact}

\sys detects four types of
\emph{impact of exploitations}
that are the building blocks of a full chain exploit:
\emph{arbitrary-chunk},
\emph{overlapping-chunk},
\emph{arbitrary-write}
and \emph{restricted-write}.
%
This approach has two benefits, namely,
expressiveness and performance.
These types are useful in
developing control-hijacking,
the ultimate goal of an attacker.
Thus, all existing techniques
lead to one of these types,
\ie can be represented by these types.
Also, it causes small performance overheads
to detect of the existence of these types
with a simple data structure
shadowing the heap space.

\CC{1} To detect arbitrary-chunk and overlapping-chunk,
\sys determines any overlapping chunks
in each allocation.
To make the check safe,
it replicates the address and size
of a chunk right after \malloc
since it could be corrupted
when a buggy action is executed.
Using the stored addresses and sizes,
it can quickly check if a chunk
overlaps with
its data structure (arbitrary-chunk)
or other chunks (overlapping-chunk).

\CC{2} To detect arbitrary-write
and restricted-write,
\sys safely replicates
its data structures,
heap containers and
global buffers,
by using a technique
known as shadow memory (see below).
During execution,
\sys synchronizes
the state of the shadow memory
whenever it performs
an action
that modifies the internal data structures:
\eg allocations for the heap container
and buffer writes for the global buffer.
At the same time,
\sys checks the divergence
between the shadowed memory
and the original memory.
Due to the explicit consistency maintained by \sys,
divergence can only arise
from the \emph{internal} operations
of a heap allocator.
Accordingly,
the divergence implies
that the executed actions
accidentally modified
\sys's data structures
via an internal operation
of the heap allocator.
For exploitation,
these actions can be reformulated
to modify other sensitive data
of an application.

\sys's fuzzing strategies
(\autoref{f:tbl-strategy})
tend to efficient detection
by limiting its analysis scope
to the data structures.
In general,
a heap exploitation technique
can corrupt any data,
leading to scanning of
the entire memory space.
However,
\sys is sufficient to check its data structure
because the only valid address from \sys's fuzzing
is either heap or its data structures.
Thus, a technique found by \sys
can only modify heap or its data structure.
\sys only cares about modification in its data structures,
but ignores one in heap
because it is hard to distinguish
with legitimate modifications (\eg by allocation)
without a deep understanding of an allocator.

\sys distinguishes arbitrary-write from restricted-write
based on the triggering heap actions.
If a divergence happens in allocation or deallocation,
it concludes restricted-write,
otherwise, arbitrary-write.
The underlying intuition
is that controlling the parameters
of the former actions
is difficult, but for the latter ones are not.
After detecting divergence,
\sys copies the original memory
to its shadow
to stop repeated detections.

\PP{Shadow memory}
\autoref{f:shadow-memory} shows
the state of the shadow memory
when executing
\autoref{f:unsafe-unlink-poc}.
\BC{1} After the first allocation,
\sys updates its heap container and corresponding shadow memory
to maintain their consistency,
which might be affected by the action.
\BC{2} It performs two more allocations
so updates the heap container and shadow memory accordingly.
\BC{3} After deallocation,
\cc{p[1]} is changed into $\star$
due to \cc{unlink()} in \ptmalloc (\autoref{f:new-unlink}).
At this point,
\sys detects divergence of the shadow memory
from the original heap container.
Since this divergence happens during deallocation,
the impact of exploitation
is limited to
\emph{restricted writes} in the heap container.
\BC{4}
In this case,
since the heap write causes the divergence,
the actions can trigger
\emph{arbitrary writes} in the heap container.
\BC{5} Since this heap write
introduces divergence in the global buffer,
the actions can lead to
\emph{arbitrary write}
in the global buffer.

\subsection{Generating PoC via Delta-Debugging}
\label{s:design:delta}

To find the root cause of exploitation,
\sys refines the test cases
by using delta-debugging~\cite{zeller:delta-debugging}.
The algorithm is simple in concept:
for each action,
\sys reevaluates the impact of exploitation
of the test cases without it.
If the impacts of the original and new test cases are equal,
then it considers the excluded action redundant
(\ie no meaningful effect to the exploitation).
The intuition behind this decision
is that
many actions are independent
(\eg buffer writes and heap writes)
so that the delta-debugging
can clearly separate non-essential actions
from the test case.
Our current algorithm is limited to
evaluating one individual action at a time.
It can be easily extended to check
the impact of a sequence or a combination of
heap actions together,
but our evaluation shows
that the current minimization scheme using single actions
is effective enough for practical uses---%
it eliminates \reducedactions of
non-essential actions on average
(see, \autoref{s:eval:delta}).

\begin{algorithm}
  \SetKwInOut{Input}{Input}
  \SetKwInOut{Output}{Output}

  \Input{$actions$ -- actions that result in an impact}

  $origImpact \leftarrow GetImpact(actions)$ \\
  $minActions \leftarrow actions$

  \For{$action \in actions$} {
    $tempActions \leftarrow minActions - action$ \\
    $tempImpact = GetImpact(tempActions)$ \\
    \If{origImpact = tempImpact} {
      $minActions \leftarrow tempActions$
    }
  }

  \Output{$minActions$ -- minimized actions that result in the same impact}

  \caption{Minimize actions that result in an impact of exploitation}
  \label{f:minimize}
\end{algorithm}

Once minimized,
it is trivial to convert the encoded test case
to a human-understandable PoC,
\eg an allocation action $\rightarrow$ \fmalloc.
We showcase the generated PoCs for newly found
exploitation primitives in~\autoref{s:appendix}.

%% file: fig/tbl-strategy.tex
\newcommand{\translinear}{$ax+b$\xspace}
\newcommand{\transoff}{$x+b$\xspace}

\begin{tabular}{clccc}
  \toprule
  \textbf{Name} & \textbf{Description} & \textbf{Align} & \textbf{Trans} & \textbf{Knowledge} \\
  \midrule
    I1 & Pre-defined constants       &                   & \\
    I2 & Offsets between pointers & \V & \transoff    & HA, BA, CA \\
    I3 & Random size (binning)       &    &              & \\
    I4 & Request size of a chunk     &    & \translinear & \\
    I5 & Chunk size of a chunk       &    & \translinear & \\
    \midrule
    P1 & NULL                        &    &              &    \\
    P2 & The buffer address          & \V & \transoff    & BA \\
    P3 & A heap address              & \V & \transoff    & HA \\
    P4 & The container address       & \V & \transoff    & CA \\
  \bottomrule
\end{tabular}

\vspace{5pt}
\textbf{I}: Integer strategy,
\textbf{P}: Pointer strategy
\textbf{HA}: Heap address
\textbf{BA}: Buffer address
\textbf{CA}: Container address

%% file: impl.tex
\section{Implementation}
\label{s:impl}

We extended American Fuzzy Lop (AFL)
to generate pseudo-random inputs
and drive our action generator
that converts the generated inputs
to heap actions.
The generator
sends a user-defined signal, SIGUSR2,
if it finds actions that result
in an impact of exploitation.
We also modified AFL
to save crashes only when it gets SIGUSR2
and ignores other signals (\eg segmentation fault),
which are not interesting in finding techniques.
We carefully implemented
the generator
not to call heap APIs implicitly
except for the pre-defined actions
for reproducing the actions.
For example,
the generator
uses the standard error
for its logging
instead of standard out,
which calls \malloc internally
for buffering.
To prevent the accidental corruption
of internal data structures,
the generator allocates
its data structures
in random addresses.
Thus,
the bug actions such as overflow
cannot modify the data structures
since they will not be adjacent
to heap chunks.
%

%% file: app.tex
\section{Case Study: Understanding Heap Exploitation}

\subsection{Discovering New Heap Exploitation Techniques}
\label{ss:new-exploit}

This section discusses
the \emph{newly discovered} exploitation techniques
against \ptmalloc.
The PoC codes are listed in~\autoref{s:appendix}.

\PP{Unsorted bin into stack (UBS)}
This technique
overwrites
the unsorted bin
to link a fake chunk
so that it can return
the address of the fake chunk
(\ie an arbitrary chunk).
This is similar to
\emph{house of lore}~\cite{malloc-des-maleficarum},
which corrupts a small bin
to achieve the same attack goal.
However,
the \emph{unsorted bin into stack} technique
requires only \emph{one} allocation,
unlike \emph{house of lore}
requires \emph{two different} allocations,
to move a chunk into a small bin list.
This technique has been added to
a community repository
that collects
heap exploitation techniques~\cite{how2heap}.

\PP{House of unsorted einherjar (HUE)}
This is a variant of \emph{house of einherjar},
which uses an off-by-one NULL byte overflow
and returns an arbitrary chunk.
In \emph{house of einherjar},
attackers should have prior knowledge
of a heap address,
\ie attackers should leak a heap pointer to break ASLR.
However, in \emph{house of unsorted einherjar},
attackers can achieve the same effect
without this pre-condition.
We named this technique,
\emph{house of unsorted einherjar},
as it interestingly combines
two techniques,
house of einherjar and unsorted bin into stack,
to relax the requirement
of the well-known exploitation technique.

\PP{Unaligned double free (UFF)}
This is a unconventional technique
that abuses
double free of a small chunk,
which is typically considered
a weak attack surface
thanks to its comprehensive security checks.
To avoid security checks,
a victim chunk for double free
should have proper metadata
and trick the next chunk under use
(\ie \previnuse $\rightarrow$ one).
%
Since the double free bug
doesn't allow arbitrary modification
of its own or the next chunk's metadata,
existing techniques
only abuse a fast bin or tcache,
which has weaker security checks than a small bin
(\eg fast-bin-dup in \autoref{f:tbl-known})
%
%

Interestingly,
\emph{unaligned double free}
bypasses these security checks
by abusing the implicit behaviors of \fmalloc.
First, it \emph{reuses}
the old metadata in a chunk
since \fmalloc does not initialize memory
by default.
Second, it fills freed space
before the next chunk
to make \previnuse of
the chunk to one.
As a result,
the technique can bypass
all security checks in \ffree,
and can successfully craft a new chunk
that overlaps with the old one.

\PP{Overlapping chunks using a small bin (OCS)}
This is a variant of overlapping-chunks (OC)
that abuses the unsorted bin
to generate an overlapping chunk
but this techniques crafts
the size of a chunk in a small bin.
Unlike OC,
it requires more actions --- three more \fmalloc and one more \ffree ---
but doesn't require attackers to control the allocation size.
When attackers cannot invoke \fmalloc with an arbitrary size,
this technique can be effective
in crafting an overlapping chunk
for exploitation.

\PP{Fast bin into other bin (FDO)}
This is another interesting technique
that allows attackers to return
an arbitrary address:
it abuses consolidation
to convert the type of a victim chunk
from the fast bin to another type.
First, it corrupts
a fast bin free list
to insert a fake chunk.
Then, it calls \consolidate
to move the fake chunk into the unsorted bin
during the deallocation process.
Unlike other techniques related to the fast bin,
this fake chunk does not have to be in the fast bin.

\subsection{Exploring Different Types of Heap Allocators}
\label{ss:other-malloc}

We also applied \sys to
two widely-used heap allocators,
\tcmalloc and \jemalloc
by Google and FreeBSD,
respectively.
Applying \sys to other allocators was trivial;
we just modified a compiler flag
to use a new allocator
other than the default, \ptmalloc.
After 24 hours of evaluation,
it found \emph{four} exploitation techniques:
three for \tcmalloc and one for \jemalloc.
We note that the number of exploitation techniques
in different implementations
does not imply their security at all.
For example,
we found only one \jemalloc exploitation technique
due to its absence of in-place metadata.
However, \jemalloc could be more vulnerable than \ptmalloc
on adjacent region overwrites.
In the following,
we discuss each techniques \sys found
(each PoC can be found in~\autoref{s:appendix}).

\PP{tcmalloc: arbitrary address return}
This technique allows an attacker
to return an arbitrary chunk in \tcmalloc.
%
It maintains a free list for each bin,
which stores its head in a static variable
and its chunks in heap.
Thus,
if we corrupt a freed chunk in heap,
allocations will
remove a chunk from list and set the list head,
finally leading to modifying the head.
Then, the next allocation will return
a corrupted memory address,
which is controlled by attackers.
Due to the similarity
between the free list of \tcmalloc
and the fast bin of \ptmalloc,
this attack is analogous to
\fastbindup in \ptmalloc.

\PP{tcmalloc: memory duplication using off-by-one}
This technique
allows attackers to trick the allocator
to return the same chunk two times
(\ie overlapping chunks)
by exploiting an off-by-one vulnerability.
It only overwrites a low byte of a freed chunk in heap
to return the same memory as before.
Unlike \ptmalloc,
which has the size right after the chunk,
\tcmalloc has the freed chunk pointer.
Therefore, the off-by-one bug
can partially overwrite
the chunk.
By overwriting the lowest byte of the pointer,
attackers can duplicate or further overlap with existing chunks.

\PP{tcmalloc, jemalloc: memory duplication using double free}
This technique allows memory duplication
in \jemalloc and \tcmalloc
by abusing a double free bug.
Unlike \ptmalloc,
since \jemalloc and \tcmalloc
do not have security checks for freed chunks,
attackers can easily trigger
traditional double free exploits
to duplicate consecutive allocations,
which are considered obsolete in \ptmalloc.

%

\begin{table}[!t]
  \centering
  \footnotesize
  \input{fig/tbl-cgc.tex}
  \caption{
    Exploitation techniques found by \sys in custom allocators of CGC.
    Except for \cc{NRFIN_00007} that implements the page heap,
    \sys successfully found exploitation techniques.
  }
  \label{f:tbl-cgc}
  \bm
\end{table}

\subsection{Evaluating Security of Custom Allocators}
\label{ss:custom-allocs}

We applied \sys to all custom heap allocators
implemented for the DARPA CGC competition---%
since many challenges share the implementation,
we selected \emph{nine} unique heap allocators
for our evaluation (see, \autoref{f:tbl-cgc}).
We implemented a missing API (\ie \cc{malloc_usable_size()})
to get the size of allocated objects,
and ran the experiment for 24 hours
for each heap allocators.

\sys found exploitation primitives for all of the tested allocators,
except for \cc{NRFIN_00007},
which places each object per page
without having any in-place metadata.
Such a page-based heap allocator looks secure
in terms of heap metadata corruption,
but it is not practical for its memory overheads,
incurring high memory usage
and causing internal fragmentation.
This experiment indicates that
the common heap designs \sys relies on
are indeed universal
in modern and custom heap allocators
(\autoref{s:background:allocators}).

One interesting thing is that
\sys found exploitation techniques
for \cc{NRFIN_00032}
that implements a heap cookie
to prevent heap overflows.
Although the cookie-based protection is not bypassible
via heap metadata corruption,
\sys found that
the implementation is vulnerable to an integer overflow.
With the integer overflow,
\sys could craft two memory blocks overlapping
without corrupting the heap cookie
by allocating an object with a proper, negative size.

Another interesting result is that
\sys automatically found
the buggy implementation
of the allocator of \cc{CROMU_00004}.
When picking a chunk for the next use,
an allocator skips a chunk
that is in-use \emph{and}
its size is less than the request size.
However, the allocator should skip a chunk
that satisfies either one of these two conditions.
Therefore,
when allocating two chunks consecutively
in which the second size is less than the first one,
the allocator responds to the second request
with the first, in-use chunk,
\ie results in overlapping chunks.
\sys successfully generated a PoC code
that exploits this buggy implementation.

\subsection{Studying Evolution of Security Features}
\label{s:eval:multiple-versions}

\begin{figure}[t]
  \centering
  \includegraphics[width=\columnwidth]{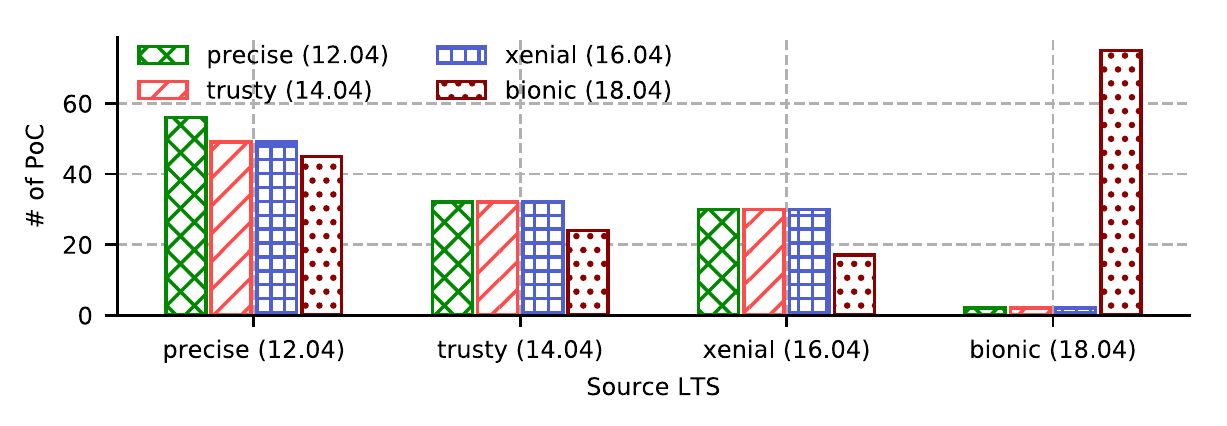}
  \vspace{-18px}
  \caption{
    The number of working PoCs from one source LTS in various Ubuntu LTS.
    For example,
    56 PoCs were generated from \precise,
    49 of them work in \trusty and \xenial,
    and 45 of them work in \bionic.
  }
  \label{f:multiple-versions}
  \bm
\end{figure}

\begin{table*}[!t]
  \begin{minipage}{.90\columnwidth}
      \centering
      \input{fig/tbl-heaphopper-new.tex}

      \caption{
        The number of discovered \emph{new} exploitation techniques
        --- the number after hash (\#) sign,
        elapsed time and corresponding models.
        Briefly, \sys discovered all four techniques,
        but \heaphopper failed to.
        We omitted FDO that has a superset model of FD,
        therefore, becomes indistinguishable to FD
        (see, \autoref{f:tbl-heaphopper-setting}).
      }
      \label{f:tbl-heaphopper-new}
  \end{minipage}
  \hspace{3px}
  \begin{minipage}{1.05\columnwidth}
      \centering
      \input{fig/tbl-heaphopper-setting.tex}
      \caption{
        Exploit-specific models for known techniques from \heaphopper.
        It is worth to note that
        results of variants
        (\ie techniques have same prerequisites, but different root causes)
        are identical for \sys
        with no specific model
        (marked with $\dagger$ and $\ddagger$ in \autoref{f:tbl-heaphopper-new} and \autoref{f:tbl-heaphopper})
        since \sys neglects the number of transactions (\ie \# Txn).
      }
      \label{f:tbl-heaphopper-setting}
      \bm
  \end{minipage}
\end{table*}

\begin{table*}[!t]
  \centering

\input{fig/tbl-heaphopper.tex}
  \caption{
    The number of discovered \emph{known} exploitation techniques
    and elapsed time for discovery in \sys and \heaphopper
    with various models.
    In summary,
    \sys outperforms \heaphopper
    with no or partly specified models,
    \eg \sys found five more techniques with no specific model (\expbase).
    Even though
    \heaphopper found one more technique than \sys
    if exploit-specific models are available,
    it suffers from false positives (marked with gray color).
    %
    %
  }
  \label{f:tbl-heaphopper}
  \bm
\end{table*}

We applied \sys 
to four versions of \ptmalloc
distributed in Ubuntu LTS:
\precise (12.04, libc 2.15), \trusty (14.04, libc 2.19),
\xenial (16.04, libc 2.23), and \bionic (18.04, libc 2.27).
In \trusty and \xenial,
a new security check, SP2,
checking the integrity of size metadata
(refer \autoref{f:tbl-security-checks}),
is backported by the Ubuntu maintainers.
To compare each version,
we perform \emph{differential} testing:
we first apply \sys to each version and generate PoCs
and then
validate the generated PoCs against other versions.
(see \autoref{f:multiple-versions}).

We identified three interesting trends
that cannot be easily obtained without \sys's automation.
First,
a new security check, in particular SP2,
successfully mitigates
a few exploitation techniques
found in an old version of \ptmalloc:
likely, the libc maintainer reacts
to a new, popular exploitation technique.
Second,
an internal design change in \bionic
rendered the most PoCs generated
from previous versions
ineffective. 
This indicates the subtleties
of the generated PoCs,
requiring precise parameters and the orders of API calls
for successful exploitation.
However,
this does not particularly mean
that a new version, \bionic, is secure;
the new component, tcache, indeed
makes exploitation much easier,
as \autoref{f:multiple-versions} shows.
Third, this new component, tcache,
which is designed to improve the performance%
~\cite{tcache-benchmark},
weakens the security of the heap allocators,
not just making it easy to attack
but also introducing new exploitation techniques.
This is similarly observed by
other researchers and communities%
~\cite{ptmalloc-fanzine, eckert:heaphopper}.


%% file: fig/tbl-cgc.tex
\newcolumntype{x}[1]{>{\centering\arraybackslash\hspace{0pt}}p{#1}}
\begin{tabular}{c x{25pt} x{25pt} x{25pt} x{25pt} }
\toprule
  \multirow{2}{*}{\textbf{Challenge}}
  & \multicolumn{4}{c}{\textbf{Impacts of exploitation}} \\
%
%
\cmidrule(lr){2-5}
              & AC & OC & AW & RW \\
\midrule
  CROMU_00003 & \V & \V & \V & \V \\
  CROMU_00004 & \V & \V & \V & \V \\
  KPRCA_00002 & \V & \V & \V & \V \\
  KPRCA_00007 & \V & \V & \V & \V \\
  NRFIN_00007 &    &    &         \\
  NRFIN_00014 & \V & \V & \V & \V \\
  NRFIN_00024 & \V & \V & \V & \V \\
  NRFIN_00027 & \V & \V & \V & \V \\
  NRFIN_00032 &    & \V &    & \V \\
\bottomrule
\end{tabular}

%% file: fig/tbl-heaphopper-new.tex
\addtolength{\tabcolsep}{-3.3pt}

\resizebox{\columnwidth}{!}{\begin{tabular}{lllll ccccc ccccc}
\multicolumn{15}{c}{\BC{1} New techniques} \\
\midrule[\heavyrulewidth]
\multirow{2}{*}{\textbf{Name}}
& \multirow{2}{*}{\textbf{Bug}}
& \multirow{2}{*}{\textbf{Impact}}
& \multirow{2}{*}{\textbf{Chunks}}
& \multirow{2}{*}{\textbf{\# Txn}}
& \multicolumn{5}{c}{\sys} & \multicolumn{5}{c}{\heaphopper} \\
\cmidrule(lr){6-10} \cmidrule(l){11-15}
&&&& &  T &  F &  O &    $\mu$ & $\sigma$ &  T &  F &  O & $\mu$ & $\sigma$ \\
\midrule
FDO &   WF &     AC &  Fast, Large &  \multicolumn{9}{c}{---} \\
\midrule
UBS &   WF &     AC &  Small &   6 & 3$^\dagger$ &  0 &  0 &   20.2m &    5m &  0 &   0 &  3 &    $\infty$ & - \\
HUE &   O1 &     AC &  Small &   9 &  2$^\ddagger$ &  0 &  1 &  14.4h &    8.9h &  0 &   0 &  3 & $\infty$ &        - \\
OCS &   OV &     OC &  Small &   9 &  3 &  0 &  0 &  17.3s &     1.2s &  0 &   0 &  3 &    $\infty$ & - \\
 UFF &   FF &     OC &  Small &   9 &  3 &  0 &  0 &  19.9s &     5.2s &  0 &   0 &  3 &    $\infty$ & - \\
\midrule
\textbf{Found}
                &&&& & 11 & 0 & 1 & \multicolumn{2}{l}{\textbf{$\Rightarrow$ \#4}}
                & 0 & 0 & 12 & \multicolumn{2}{l}{\textbf{$\Rightarrow$ \#0}}
                \\
\bottomrule
\end{tabular}}

\addtolength{\tabcolsep}{3.3pt}

\vspace{3pt}
\textbf{T}: True positives,
\textbf{F}: False positives,
\textbf{O}: Timeout,
\textbf{$\mu$}: Average time, \\
\textbf{$\sigma$}: Standard deviation of time

%% file: fig/tbl-heaphopper-setting.tex
\addtolength{\tabcolsep}{-3pt}
\resizebox{\columnwidth}{!}{
\begin{tabular}{lllllll}
\toprule
\textbf{Name} &    \textbf{Bug} &     \textbf{Impact} &   \textbf{Chunks} &  \textbf{\# Txn} &             \textbf{Size} &                    \textbf{TxnList (A list of transactions)} \\
\midrule
  FD &   WF &     AC &   Fast &   8 &            \{8\} &           M-M-F-WF-M-M \\
  UU &   O1 &  AW,RW &  Small &   6 &          \{128\} &               M-M-O1-F \\
  HS &   AF &     AC &   Fast &   4 &           \{48\} &                   AF-M \\
  PN &  O1N &     OC &  Small &  12 &  \{128,256,512\} &  M-M-M-F-O1N-M-M-F-F-M \\
  HL &   WF &     AC &  Small &   9 &     \{100,1000\} &         M-M-F-M-WF-M-M \\
  OC &   O1 &     OC &  Small &   8 &  \{120,248,376\} &           M-M-M-F-O1-M \\
  UB &   WF &  AW,RW &  Small &   7 &          \{400\} &             M-M-F-WF-M \\
  HE &   O1 &     AC &  Small &   7 &   \{56,248,512\} &             M-M-O1-F-M \\
\bottomrule
\end{tabular}}
\addtolength{\tabcolsep}{3pt}

\textbf{\# Txn}: The number of transactions, \textbf{M}: malloc, \textbf{F}: free \\

%% file: fig/tbl-heaphopper.tex
\addtolength{\tabcolsep}{-3.3pt}
\newcommand{\grayccolor}{\cellcolor{gray!40}}

\resizebox{\textwidth}{!}{\begin{tabular}{cc cccc ccccc ccccc ccccc ccccc ccccc ccccc ccccc}
& \multicolumn{30}{c}{\BC{2} Known techniques with partly specified models} & \multicolumn{10}{c}{\BC{3} Known techniques with exploit-specific models.} \\
\cmidrule[\heavyrulewidth](r){1-31} \cmidrule[\heavyrulewidth](l){32-41}
\multirow{4}{*}{\textbf{Name}} & \multicolumn{10}{c}{\textbf{\expbase}} & \multicolumn{10}{c}{\textbf{\expsize}} & \multicolumn{10}{c}{\textbf{\expactions}} & \multicolumn{10}{c}{\textbf{\expsizeactions}} \\
  \cmidrule(lr){2-11} \cmidrule(lr){12-21} \cmidrule(lr){22-31} \cmidrule(lr){32-41}
& \multicolumn{5}{c}{\sys} & \multicolumn{5}{c}{\heaphopper} & \multicolumn{5}{c}{\sys} & \multicolumn{5}{c}{\heaphopper} & \multicolumn{5}{c}{\sys} & \multicolumn{5}{c}{\heaphopper} & \multicolumn{5}{c}{\sys} & \multicolumn{5}{c}{\heaphopper} \\
 \cmidrule(lr){2-6} \cmidrule(lr){7-11} \cmidrule(lr){12-16} \cmidrule(lr){17-21} \cmidrule(lr){22-26}  \cmidrule(lr){27-31} \cmidrule(lr){32-36} \cmidrule(lr){37-41}
&  T &  F &  O &    $\mu$ & $\sigma$ &  T &  F &  O & $\mu$ & $\sigma$ &  T &  F &  O & $\mu$ & $\sigma$ &  T &  F &  O & $\mu$ & $\sigma$ &  T &  F &  O &    $\mu$ & $\sigma$ &  T &  F &  O & $\mu$ & $\sigma$ &  T &  F &  O & $\mu$ & $\sigma$ &  T &  F &  O & $\mu$ & $\sigma$ \\
\cmidrule(r){1-31} \cmidrule(l){32-41}
FD &  3 &  0 &  0 &   2.7m &     1.2m &  3 &   0 &  0 &   3.8m &     0.3s &  3 &  0 &  0 &  57.1s &    27.1s &  3 &   0 &  0 &   3.8m &     0.9s &  3 &  0 &  0 &  14.2m &     4.3m &  3 &  0 &  0 &  10.7m &     2.1m &  3 &  0 &  0 &  10.2m &     7.2m &  3 &  0 &  0 &  23.5s &     0.2s \\
UU &  3 &  0 &  0 &  57.9m &    40.4m &  0 &  0 &  3 &    $\infty$ &        - &
3   &  0 &  0 &   1.6h &     1.1h &  0 &  0 &  3 &    $\infty$ &        - &  0
&  0 &  3 &    $\infty$ &        - &  0 &  \grayccolor 3 &  0 &   3.2h &
26.3m & 0 &  0 &  3 &    $\infty$ &        - &  0 &  \grayccolor 3 &  0 &   8.2h &      13m \\
HS &  3 &  0 &  0 &   2.7m &    59.7s &  3 &   0 &  0 &  31.4s &     0.2s & 3 &  0 &  0 &   9.3m &     6.1m &  3 &   0 &  0 &  31.1s &     0.2s &  0 &  0 &  3 &    $\infty$ &        - &  3 &  0 &  0 &    56s &     0.8s &  0 &  0 & 3 &    $\infty$ &        - &  3 &  0 &  0 &  28.6s &     0.2s \\
PN &  3 &  0 &  0 &  13.3m &    24.4s &  0 &   0 &  3 &    $\infty$ & -&  3 &  0 &  0 &  16.1m &    14.9m &  0 &   0 &  3 &    $\infty$ &        - &  3 &  0 &  0 &   1.6h &      57m &  0 &  0 &  3 &    $\infty$ &        - &  3 &  0 &  0 &    26m &    12.6m &  3 &  0 &  0 &   4.3m &     1.6s \\
HL &  3$^\dagger$ &  0 &  0 &  20.2m &       5m &  0 &   0 &  3 &    $\infty$ &
- &  3 &  0 &  0 &   1.2m &    47.3s &  0 &   0 &  3 &    $\infty$ & -&  2 &  0
&  1 &  13.2h &     8.5h &  0 &  0 &  3 &    $\infty$ &        - &  3 &  0 &  0
&    21m &     9.4m &  2 &  \grayccolor 1 &  0 &   2.2m &     8.2s \\
OC &  3 &  0 &  0 &   7.1s &     5.9s &  0 &   0 &  3 &    $\infty$ & -&  3 &  0 &  0 &    20s &     5.3s &  0 &   0 &  3 &    $\infty$ &        - &  3 &  0 &  0 &     6s &     2.4s &  3 &  0 &  0 &  22.1h &    33.2m &  3 &  0 &  0 &  26.6s &      34s &  3 &  0 &  0 &   3.2m &       2s \\
UB &  3 &  0 &  0 &  36.8s &    22.8s &  3 &   0 &  0 &  21.8s &     0.2s &  3 &  0 &  0 &   4.7s &     3.1s &  3 &   0 &  0 &  21.9s &     0.3s &  3 &  0 &  0 &  24.8s &    14.9s &  3 &  0 &  0 &  47.6s &     0.3s &  3 &  0 &  0 &  12.6s &     9.5s &  3 &  0 &  0 &  19.5s &     0.7s \\
HE &  2$^\ddagger$ &  0 &  1 &  14.4h &     8.9h &  0 &   0 &  3 & $\infty$ &
- &  2 &  0 &  1 &   9.3h &    10.4h &  0 &   0 &  3 & $\infty$ &        - &  0
&  0 &  3 &    $\infty$ &        - &  0 &  0 &  3 & $\infty$ &        - &  0 &
0 &  3 &    $\infty$ &        - &  0 &  \grayccolor 3 &  0 &   6.8m &     6.4s \\
\cmidrule(r){1-31} \cmidrule(l){32-41}
\textbf{Found}  
                & 23 & 0 & 1 & \multicolumn{2}{l}{\textbf{$\Rightarrow$ \#8}}
                & 9 & 0 & 15 & \multicolumn{2}{l}{\textbf{$\Rightarrow$ \#3}}
                & 23 & 0 & 1 & \multicolumn{2}{l}{\textbf{$\Rightarrow$ \#8}}
                & 9 & 0 & 15 & \multicolumn{2}{l}{\textbf{$\Rightarrow$ \#3}}
                & 14 & 0 & 10 &\multicolumn{2}{l}{\textbf{$\Rightarrow$ \#5}}
                & 12 & 3 & 9 & \multicolumn{2}{l}{\textbf{$\Rightarrow$ \#4}}
                & 15 & 0 & 9 & \multicolumn{2}{l}{\textbf{$\Rightarrow$ \#5}}
                & 17 & 7 & 0 & \multicolumn{2}{l}{\textbf{$\Rightarrow$ \#6}} \\ 
\cmidrule[\heavyrulewidth](r){1-31} \cmidrule[\heavyrulewidth](l){32-41}
\end{tabular}}

\addtolength{\tabcolsep}{3.3pt}

%% file: eval.tex
\section{Evaluation}
\label{s:eval}

This section tries to answer the following questions:

\begin{enumerate}
\item How effective is \sys
  in finding exploitation techniques
  compared to the state-of-the-art technique?
\item How exhaustively can \sys explore
  the security-critical state space?
\item How effective is delta-debugging
  in minimizing heap actions?
\end{enumerate}

\PP{Evaluation setup}
We conducted all the experiments
on Intel Xeon E7-4820 with 256~GB RAM.
We used 256 random bytes as a seed
that is used to indicate
a starting point of the state exploration,
and ran each experiment three times
(24 $\times$ 3 hours)
to reduce statistical variance.
Selecting a seed in \sys is not critical
in discovering new exploit techniques
as it tends to converge
during the state exploration.

\subsection{Comparison to HeapHopper}

\heaphopper~\cite{eckert:heaphopper}
was recently proposed
to analyze exploitation techniques
by using symbolic execution.
To overcome the state explosion in symbolic execution,
\heaphopper tightly
encodes the prior knowledge of exploit techniques
into its models,
e.g., the number of \emph{transactions}
(\ie equivalent to non-write actions in \sys),
allocation sizes
(\ie guiding the use of specific bins),
and even a certain order of transactions
(\ie a sequence of non-write actions).
By relying on this model,
it could incrementally perform
the symbolic execution
for all permutations of transactions in order.
Unfortunately,
its key idea---guiding the state exploration with detailed models---
limits its capability
only to validate
\emph{known} exploitation techniques,
unlike our approach can find
\emph{unknown} techniques
with \emph{fuzzing}.

We performed three experiments
that objectively compare both approaches:
\BC{1} finding \emph{unknown} techniques with no exploit-specific model
(\ie applying \heaphopper to \sys's task),
\BC{2} finding \emph{known} techniques with partly specified models
(\ie evaluating the roles of specified models in each approach),
and \BC{3} finding \emph{known} techniques with exploit-specific models
(\ie applying \sys to \heaphopper's task).
In the experiments,
we considered variants of exploit techniques%
\footnote{Exploit techniques often have the same prerequisite
but different root causes such as UBS and HL.}
as a equal class
since both systems cannot distinguish
their subtle differences.
We ran each experiment
three times with 24-hour timeout
for proper statistical comparison%
~\cite{klees:fuzzeval}.

\vspace{2px}
\noindent\textbf{\BC{1} New techniques.}
We first check if \heaphopper's approach can be used
to find previously \emph{unknown} exploitation techniques
that \sys found (see, \autoref{ss:new-exploit}).
To apply \heaphopper,
we provided relaxed models
that specify all boundary sizes for all bins
but limit the number of transactions
following our PoCs
as shown in \autoref{f:tbl-heaphopper-new}.
Note that,
in theory,
such relaxation is general enough
to discover new techniques
given \emph{infinite} amount of computing resources.
In the experiment,
FDO is excluded
because its model is a superset of FD;
having FDO simply makes \sys and \heaphopper converged to FD.

\heaphopper fails to identify
all \emph{unknown} exploitation primitives
with no exploit-specific models
(see \autoref{f:tbl-heaphopper-new}).
In fact,
it encounters a few fundamental problems of symbolic execution:
1) exponentially growing permutations of transactions
and 2) huge search spaces
in selecting proper size parameters
to trigger exploitation.
Although
\heaphopper demonstrated
a successful state exploration
of seven transactions with three size parameters
(\S7.1 in \cite{eckert:heaphopper}),
the search space required for discovering \emph{unknown} techniques
are much larger,
rendering \heaphopper's approach
computationally infeasible.
On the contrary,
\sys successfully explores the search space
by using the random strategies,
and indeed discovers unknown techniques,
showing the practicality of our approach.

\vspace{2pt}
\noindent\textbf{\BC{2} Known techniques with partly specified models.}
We evaluated how specified, exploit-specific models
play a role for both approaches.
In particular,
we tested both systems with
the exploit-specific models,
namely,
the size parameters (\expsize)
and a sequence of transactions (\expactions),
used in \heaphopper
(see, \autoref{f:tbl-heaphopper-setting}).
To prevent each systems from converging to easy-to-find techniques,
we tested each model on
top of the baseline heap model
(\ie \expbase).

This experiment
(\ie \BC{2} in \autoref{f:tbl-heaphopper}) shows
that \sys performs better than \heaphopper
with no or partly specified models:
\sys found five more \emph{known} techniques
than \heaphopper
in \expbase and \expsize.
We observed two interesting behaviors of \sys.
First, when additional information is provided (\ie guided),
we expected that \sys would detect target exploitation techniques quicker
(\eg 20.2$\rightarrow$1.2m in HL with \expsize),
but often slower (\eg 2.7$\rightarrow$9.3m in HS with \expsize).
Such behaviors indicate that
additional information tends to misguide \sys;
perhaps, forcing it to explore the unlikely state space.
Second,
unlike \sys,
\heaphopper behaves better
with exploit-specific models:
finding one more techniques
when \expactions is provided with \expbase.
This result shows that
a precise model plays an essential role
in symbolic execution.
In brief,
\sys's is particularly preferable
when exploring \emph{unknown} search space,
but similarly effective
when exploring with the partly specified model.

\vspace{2pt}
\noindent\textbf{\BC{3} Known techniques with exploit-specific models}
When both \expsize and \expactions are provided,
\heaphopper's approach works better:
it found one more \emph{known} technique 
and found four techniques quicker than \sys
(as illustrated in \BC{3} in \autoref{f:tbl-heaphopper}).
This indicates the importance of accurate model
for the symbolic execution
in effectively reducing the search space.
We observed one interesting behavior of \heaphopper
in this experiment.
With more exploit models specified,
\heaphopper tends to suffer
from false positives
(\ie incorrectly claiming the discovery of exploit techniques)
because of its internal complexity---%
we confirmed these false positives
with \heaphopper's authors.
In specific,
\heaphopper failed to find UU and UE
because of the complicated analysis
of underlying framework,
\cc{angr}~\cite{shoshitaishvili:angr},
as noted in the paper~\cite{eckert:heaphopper}.
On the contrary,
\sys's approach does not introduce false positive
thanks to its comprehensive detection method
using shadow memory.

This experiment also highlights
an interesting design decision of \sys:
separating the exploration and reducing phases (\ie minimization).
With no exploit-specific guidance,
\sys can freely explore the search space,
and so increase the probability
of satisfying the precondition
of certain exploitation techniques.
For example,
if the sequence of transactions of UU (M-M-O1-F) is enforced,
\sys should craft a fake chunk
within a relatively small period
(\ie between four actions)
to trigger the exploit;
otherwise, 
\sys has a higher probability
to formulate a fake chunk
by executing more, perhaps redundant actions.
However, such redundancy is acceptable in \sys
thanks to our minimization phase
that effectively reduce inessential actions
from the found exploit.

We also confirmed that
\sys can find all tcache-related techniques%
~\cite{ptmalloc-fanzine}
and \emph{house-of-force},
which \heaphopper fails to find,
because an arbitrary size allocation is required.
\sys can find these techniques
within a few minutes
as they require less than five transactions.

%
\subsection{Security Check Coverage}
\label{s:eval:abort}

To show how exhaustively \sys explore
the security-sensitive part of the state space,
we counted the number of security checks 
executed by \sys.
In 24 hours of exploration,
\sys executed 18 out of 21 security checks of \ptmalloc:
it failed to cover D2, D4 and SP5 in~\autoref{f:tbl-security-checks}.
We note that SP5 is related to
a concurrency bug,
which is outside of the scope of this work.
D1 and D4 require a strict relationship
between large chunks
(\eg the sizes of two chunks are not equal but less than the minimum size),
which is probably too stringent
for any randomization-based strategies.

\subsection{Delta-Debugging-Based Minimization}
\label{s:eval:delta}

The minimization techniques
based on delta-debugging
is effective in simplifying the generated PoCs
for further analysis.
It effectively reduces \reducedactions
redundant actions from original PoCs
(refer \autoref{s:eval:multiple-versions})
and emits small PoCs
that contain 26.1 lines on average
(see, \autoref{f:tbl-delta}).
Although our minimization is preliminary
(\ie eliminating one independent action per testing),
the final PoC is sufficiently small 
for manual analysis.

\begin{table}[!t]
  \centering
  \footnotesize
  \input{fig/tbl-delta.tex}

  \caption{
    Average and standard derivation of lines of
    raw and minimized PoCs using delta debugging.
    It shows that the delta debugging
    successfully removes \reducedactions
    of redundant actions.
  }
  \label{f:tbl-delta}
  \bm
\end{table}

%% file: fig/tbl-delta.tex
\begin{tabular}{crrrr}
  \toprule
  \multirow{2}{*}{\textbf{Version}}
    & \multicolumn{2}{c}{\textbf{Raw}}
    & \multicolumn{2}{c}{\textbf{Minimized}} \\
\cmidrule(lr){2-5}
          & Mean & Std. dev & Mean & Std. dev \\
  \midrule
  2.15    & 112.6 & 161   & 25.9 (-77.0 \%)    & 25.3 \\
    2.19    & 110.8 & 145   & 23.3 (-79.0 \%)  &  4.6 \\
    2.23    &  98.3 & 120 & 22.5 (-77.1 \%)  &  6.2 \\
    2.27    & 344.2 & 177   & 33   (-90.4 \%)   &  8.8 \\
  \midrule
    Average & 166.5 & 150.8 & 26.2 (-84.3 \%) & 11.2 \\
  \bottomrule
\end{tabular}

%% file: discuss.tex
\section{Discussion and Limitations}
\label{s:discuss}

\PP{Completeness}
\sys is fundamentally \emph{incomplete}
due to its random nature,
so it is not at all surprising
if someone discover other heap exploitation techniques.
\heaphopper,
on the other hand,
is \emph{complete} in terms of
\emph{given models}:
\ie exploring all combinations of transactions given the length of transactions.
Since their models are incomplete (or often error-prone),
the proper use of each approach
is dependent on
the target use cases.
For example,
if one is looking for a practical solution,
\sys would be a more preferable platform to start with.

\PP{Overfitting to fuzzing strategies}
\sys builds upon an intuition in which
modern heap allocators
follow common design idioms:
binning, in-place metadata and cardinal data.
This helps \sys in reducing its search space
but might cause an overfitting problem:
the discovered exploit techniques are
too specific to certain designs of heap allocators.
To apply \sys to non-conventional implementation
of heap allocators,
one might have to devise
own models for proper space reduction.

%

\PP{Enhancing mitigation}
\sys can also be used
to improve the countermeasures
of heap allocators.
For example,
the generated PoCs for each security check
(\autoref{s:eval:abort})
can be used for unit testing,
and the PoCs for each exploit techniques
(\autoref{s:eval:multiple-versions})
can be used for regression test cases.
Also, thanks to \sys's automatic nature,
developers can use it to find
potential security issues of own changes:
\eg \sys would automatically mark tcache
a red flag
if run before the release.



%% file: relwk.tex
\section{Related work}
\label{s:relwk}

\PP{Automatic exploit generation (AEG)}
Automatic discovery of heap exploit techniques
is a small step toward
AEG's ambitious vision~\cite{brumley:apeg, avgerinos:aeg},
but it is worth emphasizing
its importance and difficulty.
Despite several attempts to accomplish
fully automated exploit generation%
~\cite{brumley:apeg, avgerinos:aeg,
cha:mayhem, schwartz:q, lu:use-before-init,
wang:revery, heelan:automatic, repel:modular},
AEG, particularly for heap vulnerabilities,
is so sophisticated and difficult
that all the state-of-the-art cyber reasoning systems from DARPA CGC,
(\ie systems finding and exploiting vulnerabilities automatically%
~\cite{forallsecure:cgc, shellphish:cgc,
grammatech:cgc, trailofbits:cgc}),
failed to address;
according to organizers, only a single heap
vulnerability was successfully exploited
in the CGC final event.
Recently,
Repel \etal~\cite{repel:modular}
proposes a symbolic-execution-based approach
aiming at AEG for heap vulnerabilities,
but only works for old allocators
without security checks.
Heelan \etal~\cite{heelan:automatic} demonstrates
an automatic method to find an object layout for exploitation
specific to an application.
Unlike the prior work,
\sys focuses on finding
heap exploitation techniques,
which are re-usable across applications,
in modern allocators
with full security checks.
%
%
%
%
%
%

\PP{Fuzzing beyond crashes}
There has been a large body of attempts
to extend fuzzing
to find bugs beyond memory safety~\cite{afl, syzkaller}.
They often use a differential testing,
which we used for minimization,
to find semantic bugs:
\eg compilers~\cite{yang:csmith},
cryptographic libraries~\cite{petsios:nezha, brubaker:frankencerts},
JVM implementations~\cite{chen:jvm}
and learning systems~\cite{pei:deepxplore}.
Recently,
SlowFuzz~\cite{petsios:slowfuzz}
uses fuzzing to find algorithmic complexity bugs,
and IMF~\cite{wang:imf} to spot
similar code in binary.
%

\PP{Application-aware fuzzing}
Application-aware fuzzing is one of the attempts to
reduce the search space of fuzzing.
In this regard,
there have been attempts to use
static and dynamic analysis%
~\cite{peng:t-fuzz, rawat:vuzzer, li:steelix, chen:angora},
bug descriptions~\cite{you:semfuzz},
and real-world applications%
~\cite{chen:iotfuzzer, kim:cab-fuzz, han:imf}
to extract target-specific information for fuzzing.
Moreover,
to reduce the search space
for applications
that require well-formed inputs,
researchers have embedded
domain-specific knowledge
such as
grammar~\cite{holler:lang-fuzz, yang:csmith, wang:skyfire}
or structure~\cite{petsios:nezha, brubaker:frankencerts}
in their fuzzing.
Similar to these works,
\sys reduces its search space
by considering its targets and
memory allocators,
particularly
exploiting their common designs.

%% file: conclusion.tex
\section{Conclusion}
\label{s:conclusion}

In this paper,
we present \sys,
a new approach using fuzzing
to automatically discover
new heap exploitation techniques.
Two key enablers of \sys's approach
are to reduce the search space of fuzzing
by abstracting the common design
of modern heap allocators,
and to devise a method to quickly
estimate the possibility of heap exploitation.
Our evaluation with three real-world and a few custom heap allocators
shows that
\sys's approach can
effectively formulate new exploitation primitives
regardless of their underlying implementation.

%% file: appendix.tex
\appendix
\renewcommand\thefigure{\thesection.\arabic{figure}}
\setcounter{figure}{0}
\section{Appendix}
\label{s:appendix}

\begin{figure}[!ht]
  \begin{subfigure}{\columnwidth}
    \input{code/tcmalloc-arbitrary-address-return.c}
    \caption{
      An exploitation technique
      for \tcmalloc returning an arbitrary address
      that was found by \sys.
    }
    \label{f:tcmalloc-arbitrary-address-return}
    \vspace{5px}
  \end{subfigure}
  \begin{subfigure}{\columnwidth}
    \input{code/tcmalloc-duplicate.c}
    \caption{
      An exploitation technique for \tcmalloc
      returning duplicate addresses
      using off-by-one bug
      that was found by \sys.
    }
    \label{f:tcmalloc-duplicate}
    \vspace{5px}
  \end{subfigure}
  \begin{subfigure}{\columnwidth}
    \input{code/jemalloc-double-free.c}
    \caption{
      An exploitation technique
      for \tcmalloc and \jemalloc triggering double free
      that was found by \sys.
    }
    \label{f:jemalloc-double-free}
  \end{subfigure}

  \caption{Exploitation techniques
  found by \sys in \tcmalloc and \jemalloc}
  \label{f:other-allocator-poc}
\end{figure}

\begin{figure}[!t]
  \input{code/unsorted-bin-into-stack.c}
  \caption{
    A new exploitation technique that \sys found,
    named \emph{unsorted bin into stack},
    which returns arbitrary memory
    by corrupting the unsorted bin.
  }
  \label{f:unsorted-bin-into-stack}
\end{figure}

\begin{figure}[!t]
  \input{code/fast-bin-to-other-bin.c}
  \caption{
    A new exploitation technique that \sys found,
    named \emph{fast bin into other bin},
    which returns arbitrary memory of non-fast bin size.
  }
  \label{f:fast-bin-to-other-bin}
\end{figure}

\begin{figure}[!t]
  \input{code/overlapping-chunks-smallbin.c}
  \caption{
    A new exploitation technique that \sys found,
    named \emph{overlapping chunks smallbin},
    which returns overlapped chunk in small bin.
    Even though this requires
    more steps than overlapping chunks,
    it does not need \emph{accurate} size for allocation.
  }
  \label{f:overlapping-chunks-smallbin}
\end{figure}

\begin{figure}[!t]
  \input{code/unaligned-double-free.c}
  \caption{
    A new exploitation technique that \sys found,
    named \emph{unaligned double free},
    which returns overlapped chunks
    by the double free bug.
  }
  \label{f:unaligned-double-free}
\end{figure}

\begin{figure}[!t]
  \input{code/house-of-unsorted-einherjar.c}
  \caption{
    A new exploitation technique that \sys found, named house of unsorted einherjar.
    This is variant of a known heap exploitation technique, house of einherjar,
    but it does not require a heap address, which is essential for the old technique.
  }
  \label{f:house-of-unsorted-einherjar}
\end{figure}

%% file: code/tcmalloc-arbitrary-address-return.c.tex
\begin{Verbatim}[commandchars=\\\{\},codes={\catcode`\$=3\catcode`\^=7\catcode`\_=8}]
  \PY{c+c1}{// [PRE\PYZhy{}CONDITION]}
  \PY{c+c1}{//    sz : any size}
  \PY{c+c1}{// [BUG] buffer overflow}
  \PY{c+c1}{// [POST\PYZhy{}CONDITION]}
  \PY{c+c1}{//    malloc(sz) == dst}
  \PY{k+kt}{void}\PY{o}{*} \PY{n}{p} \PY{o}{=} \PY{n}{malloc}\PY{p}{(}\PY{n}{sz}\PY{p}{)}\PY{p}{;}
  \PY{c+c1}{// [BUG] overflowing p}
  \PY{c+c1}{// tcmalloc has a next chunk address at the end of a chunk}
  \PY{o}{*}\PY{p}{(}\PY{k+kt}{void}\PY{o}{*}\PY{o}{*}\PY{p}{)}\PY{p}{(}\PY{n}{p} \PY{o}{+} \PY{n}{malloc\PYZus{}usable\PYZus{}size}\PY{p}{(}\PY{n}{p}\PY{p}{)}\PY{p}{)} \PY{o}{=} \PY{n}{dst}\PY{p}{;}

  \PY{c+c1}{// this malloc changes a next chunk address into dst}
  \PY{n}{malloc}\PY{p}{(}\PY{n}{sz}\PY{p}{)}\PY{p}{;}

  \PY{n}{assert}\PY{p}{(}\PY{n}{malloc}\PY{p}{(}\PY{n}{sz}\PY{p}{)} \PY{o}{=}\PY{o}{=} \PY{n}{dst}\PY{p}{)}\PY{p}{;}
\end{Verbatim}

%% file: code/tcmalloc-duplicate.c.tex
\begin{Verbatim}[commandchars=\\\{\},codes={\catcode`\$=3\catcode`\^=7\catcode`\_=8}]
  \PY{c+c1}{// [PRE\PYZhy{}CONDITION]}
  \PY{c+c1}{//    sz : any size \PYZlt{} 0x100}
  \PY{c+c1}{// [BUG] off\PYZhy{}by\PYZhy{}one null overflow}
  \PY{c+c1}{// [POST\PYZhy{}CONDITION]}
  \PY{c+c1}{//    malloc(sz) == already allocated one}
  \PY{k+kt}{void}\PY{o}{*} \PY{n}{p} \PY{o}{=} \PY{n}{malloc}\PY{p}{(}\PY{n}{sz}\PY{p}{)}\PY{p}{;}

  \PY{c+c1}{// p\PYZsq{}s lowest byte is zero in tcmalloc}
  \PY{n}{assert}\PY{p}{(}\PY{p}{(}\PY{k+kt}{intptr\PYZus{}t}\PY{p}{)}\PY{n}{p} \PY{o}{\PYZam{}} \PY{l+m+mh}{0xff} \PY{o}{=}\PY{o}{=} \PY{l+m+mi}{0}\PY{p}{)}\PY{p}{;}

  \PY{c+c1}{// [BUG] off\PYZhy{}by\PYZhy{}one\PYZhy{}null overflow}
  \PY{c+c1}{// it clears a lowest byte of a next chunk address}
  \PY{c+c1}{// to make the next chunk same with \PYZsq{}p\PYZsq{}}
  \PY{o}{*}\PY{p}{(}\PY{k+kt}{char}\PY{o}{*}\PY{p}{)}\PY{p}{(}\PY{n}{p} \PY{o}{+} \PY{n}{malloc\PYZus{}usable\PYZus{}size}\PY{p}{(}\PY{n}{p}\PY{p}{)}\PY{p}{)} \PY{o}{=} \PY{l+m+mi}{0}\PY{p}{;}

  \PY{c+c1}{// it updates the next chunk address == \PYZsq{}p\PYZsq{}}
  \PY{n}{malloc}\PY{p}{(}\PY{n}{sz}\PY{p}{)}\PY{p}{;}

  \PY{n}{assert}\PY{p}{(}\PY{n}{p} \PY{o}{=}\PY{o}{=} \PY{n}{malloc}\PY{p}{(}\PY{n}{sz}\PY{p}{)}\PY{p}{)}\PY{p}{;}
\end{Verbatim}

%% file: code/jemalloc-double-free.c.tex
\begin{Verbatim}[commandchars=\\\{\},codes={\catcode`\$=3\catcode`\^=7\catcode`\_=8}]
  \PY{c+c1}{// [PRE\PYZhy{}CONDITION]}
  \PY{c+c1}{//    sz : any size}
  \PY{c+c1}{// [BUG] double free}
  \PY{c+c1}{// [POST\PYZhy{}CONDITION]}
  \PY{c+c1}{//    malloc(sz) == malloc(sz)}
  \PY{k+kt}{void}\PY{o}{*} \PY{n}{p} \PY{o}{=} \PY{n}{malloc}\PY{p}{(}\PY{n}{sz}\PY{p}{)}\PY{p}{;}
  \PY{n}{free}\PY{p}{(}\PY{n}{p}\PY{p}{)}\PY{p}{;}

  \PY{c+c1}{// [BUG] free \PYZsq{}p\PYZsq{} again}
  \PY{c+c1}{// this is allowed due to lack of security checks}
  \PY{n}{free}\PY{p}{(}\PY{n}{p}\PY{p}{)}\PY{p}{;}

  \PY{n}{assert}\PY{p}{(}\PY{n}{malloc}\PY{p}{(}\PY{n}{sz}\PY{p}{)} \PY{o}{=}\PY{o}{=} \PY{n}{malloc}\PY{p}{(}\PY{n}{sz}\PY{p}{)}\PY{p}{)}\PY{p}{;}
\end{Verbatim}

%% file: code/unsorted-bin-into-stack.c.tex
\begin{Verbatim}[commandchars=\\\{\},codes={\catcode`\$=3\catcode`\^=7\catcode`\_=8}]
  \PY{c+c1}{// [PRE\PYZhy{}CONDITION]}
  \PY{c+c1}{//    sz : any non\PYZhy{}fast\PYZhy{}bin size}
  \PY{c+c1}{// [BUG] buffer overflow}
  \PY{c+c1}{// [POST\PYZhy{}CONDITION]}
  \PY{c+c1}{//    malloc(sz) = dst + offsetof(struct malloc\PYZus{}chunk, fd)}
  \PY{k+kt}{void}\PY{o}{*} \PY{n}{p1} \PY{o}{=} \PY{n}{malloc}\PY{p}{(}\PY{n}{sz}\PY{p}{)}\PY{p}{;}
  \PY{k+kt}{void}\PY{o}{*} \PY{n}{p2} \PY{o}{=} \PY{n}{malloc}\PY{p}{(}\PY{n}{sz}\PY{p}{)}\PY{p}{;}
  \PY{k+kt}{void}\PY{o}{*} \PY{n}{p3} \PY{o}{=} \PY{n}{malloc}\PY{p}{(}\PY{n}{sz}\PY{p}{)}\PY{p}{;}

  \PY{c+c1}{// move p2 to the unsorted bin}
  \PY{n}{free}\PY{p}{(}\PY{n}{p2}\PY{p}{)}\PY{p}{;}

  \PY{c+c1}{// create a fake chunk at dst}
  \PY{k}{struct} \PY{n}{malloc\PYZus{}chunk} \PY{o}{*}\PY{n}{fake} \PY{o}{=} \PY{n}{dst}\PY{p}{;}
  \PY{c+c1}{// set fake\PYZhy{}\PYZgt{}size to be the chunk size of the last allocation}
  \PY{n}{fake}\PY{o}{\PYZhy{}}\PY{o}{\PYZgt{}}\PY{n}{size} \PY{o}{=} \PY{n}{chunk\PYZus{}size}\PY{p}{(}\PY{n}{sz}\PY{p}{)}\PY{p}{;}
  \PY{c+c1}{// set fake\PYZhy{}\PYZgt{}bk to any writable address to avoid a crash}
  \PY{n}{fake}\PY{o}{\PYZhy{}}\PY{o}{\PYZgt{}}\PY{n}{bk} \PY{o}{=} \PY{n}{fake}\PY{p}{;}

  \PY{c+c1}{// [BUG] overflowing p1}
  \PY{k}{struct} \PY{n}{malloc\PYZus{}chunk} \PY{o}{*}\PY{n}{c2} \PY{o}{=} \PY{n}{raw\PYZus{}to\PYZus{}chunk}\PY{p}{(}\PY{n}{p2}\PY{p}{)}\PY{p}{;}
  \PY{c+c1}{// size should be smaller than the next allocation size}
  \PY{c+c1}{// to avoid returning c2 in the next allocation}
  \PY{c+c1}{// size shouldn\PYZsq{}t be too small due to a security check}
  \PY{n}{c2}\PY{o}{\PYZhy{}}\PY{o}{\PYZgt{}}\PY{n}{size} \PY{o}{=} \PY{l+m+mi}{2} \PY{o}{*} \PY{k}{sizeof}\PY{p}{(}\PY{k+kt}{size\PYZus{}t}\PY{p}{)}\PY{p}{;}
  \PY{c+c1}{// set the next pointer in the unsorted bin}
  \PY{n}{c2}\PY{o}{\PYZhy{}}\PY{o}{\PYZgt{}}\PY{n}{bk} \PY{o}{=} \PY{n}{fake}\PY{p}{;}

  \PY{c+c1}{// now unsorted bin: c2 \PYZhy{}\PYZgt{} fake,}
  \PY{c+c1}{// and c2 is too small for the request.}
  \PY{c+c1}{// therefore, next allocation returns the fake chunk}
  \PY{n}{assert}\PY{p}{(}\PY{n}{malloc}\PY{p}{(}\PY{n}{sz}\PY{p}{)} \PY{o}{=}\PY{o}{=} \PY{n}{fake} \PY{o}{+} \PY{n}{offsetof}\PY{p}{(}\PY{k}{struct} \PY{n}{malloc\PYZus{}chunk}\PY{p}{,} \PY{n}{fd}\PY{p}{)}\PY{p}{)}\PY{p}{;}
\end{Verbatim}

%% file: code/fast-bin-to-other-bin.c.tex
\begin{Verbatim}[commandchars=\\\{\},codes={\catcode`\$=3\catcode`\^=7\catcode`\_=8}]
  \PY{c+c1}{// [PRE\PYZhy{}CONDITION]}
  \PY{c+c1}{//    fsz: any fast bin size}
  \PY{c+c1}{//    sz: any non\PYZhy{}fast\PYZhy{}bin size}
  \PY{c+c1}{//    lsz: any largebin size}
  \PY{c+c1}{// [BUG] write free memory}
  \PY{c+c1}{// [POST\PYZhy{}CONDITION]}
  \PY{c+c1}{//    malloc(sz) = fake \PYZhy{} offsetof(struct malloc\PYZus{}chunk, fd)}
  \PY{k+kt}{void}\PY{o}{*} \PY{n}{p1} \PY{o}{=} \PY{n}{malloc}\PY{p}{(}\PY{n}{fsz}\PY{p}{)}\PY{p}{;}
  \PY{n}{free}\PY{p}{(}\PY{n}{p1}\PY{p}{)}\PY{p}{;}

  \PY{c+c1}{// create a fake chunk}
  \PY{k}{struct} \PY{n}{malloc\PYZus{}chunk} \PY{o}{*}\PY{n}{fake} \PY{o}{=} \PY{n}{dst}\PY{p}{;}
  \PY{c+c1}{// set P=1 to avoid a security check}
  \PY{n}{fake}\PY{o}{\PYZhy{}}\PY{o}{\PYZgt{}}\PY{n}{size} \PY{o}{=} \PY{n}{chunk\PYZus{}size}\PY{p}{(}\PY{n}{sz}\PY{p}{)} \PY{o}{|} \PY{l+m+mi}{1}\PY{p}{;}
  \PY{n}{fake}\PY{o}{\PYZhy{}}\PY{o}{\PYZgt{}}\PY{n}{fd} \PY{o}{=} \PY{n+nb}{NULL}\PY{p}{;}

  \PY{c+c1}{// create \PYZsq{}fake2\PYZsq{}: a next chunk of \PYZsq{}fake\PYZsq{}}
  \PY{k}{struct} \PY{n}{malloc\PYZus{}chunk} \PY{o}{*}\PY{n}{fake2} \PY{o}{=} \PY{n}{dst} \PY{o}{+} \PY{n}{chunk\PYZus{}size}\PY{p}{(}\PY{n}{sz}\PY{p}{)}\PY{p}{;}
  \PY{c+c1}{// set P=1 to avoid a security check}
  \PY{n}{fake2}\PY{o}{\PYZhy{}}\PY{o}{\PYZgt{}}\PY{n}{size} \PY{o}{=} \PY{l+m+mi}{1}\PY{p}{;}

  \PY{k}{struct} \PY{n}{malloc\PYZus{}chunk} \PY{o}{*}\PY{n}{c1} \PY{o}{=} \PY{n}{raw\PYZus{}to\PYZus{}chunk}\PY{p}{(}\PY{n}{p1}\PY{p}{)}\PY{p}{;}
  \PY{c+c1}{// [BUG] set a forward pointer of fast bin into fake}
  \PY{c+c1}{// this can be done by a normal heap write since p4 is allocated}
  \PY{n}{c1}\PY{o}{\PYZhy{}}\PY{o}{\PYZgt{}}\PY{n}{fd} \PY{o}{=} \PY{n}{fake}\PY{p}{;}

  \PY{c+c1}{// now a fast bin list: c1 \PYZhy{}\PYZgt{} fake}
  \PY{c+c1}{// call malloc\PYZus{}consolidate to move}
  \PY{c+c1}{// \PYZsq{}fake\PYZsq{} to the unsorted bin}
  \PY{n}{malloc}\PY{p}{(}\PY{n}{lsz}\PY{p}{)}\PY{p}{;}

  \PY{n}{assert}\PY{p}{(}\PY{n}{raw\PYZus{}to\PYZus{}chunk}\PY{p}{(}\PY{n}{malloc}\PY{p}{(}\PY{n}{sz}\PY{p}{)}\PY{p}{)} \PY{o}{=}\PY{o}{=} \PY{n}{fake}\PY{p}{)}\PY{p}{;}
\end{Verbatim}

%% file: code/overlapping-chunks-smallbin.c.tex
\begin{Verbatim}[commandchars=\\\{\},codes={\catcode`\$=3\catcode`\^=7\catcode`\_=8}]
  \PY{c+c1}{// [PRE\PYZhy{}CONDITION]}
  \PY{c+c1}{//    sz : any small bin size}
  \PY{c+c1}{//    sz2 : any small bin size}
  \PY{c+c1}{//    assert(sz2 \PYZgt{} sz)}
  \PY{c+c1}{// [BUG] buffer overflow}
  \PY{c+c1}{// [POST\PYZhy{}CONDITION]}
  \PY{c+c1}{//    two chunks overlap}

  \PY{k+kt}{void}\PY{o}{*} \PY{n}{p1} \PY{o}{=} \PY{n}{malloc}\PY{p}{(}\PY{n}{sz}\PY{p}{)}\PY{p}{;}
  \PY{k+kt}{void}\PY{o}{*} \PY{n}{p2} \PY{o}{=} \PY{n}{malloc}\PY{p}{(}\PY{n}{sz}\PY{p}{)}\PY{p}{;}
  \PY{k+kt}{void}\PY{o}{*} \PY{n}{p3} \PY{o}{=} \PY{n}{malloc}\PY{p}{(}\PY{n}{sz}\PY{p}{)}\PY{p}{;}

  \PY{c+c1}{// move p2 to the unsorted bin}
  \PY{n}{free}\PY{p}{(}\PY{n}{p2}\PY{p}{)}\PY{p}{;}

  \PY{c+c1}{// move p2 to the small bin}
  \PY{k+kt}{void}\PY{o}{*} \PY{n}{p4} \PY{o}{=} \PY{n}{malloc}\PY{p}{(}\PY{n}{sz2}\PY{p}{)}\PY{p}{;}

  \PY{c+c1}{// [BUG] overflowing p1}
  \PY{k}{struct} \PY{n}{malloc\PYZus{}chunk} \PY{o}{*}\PY{n}{c2} \PY{o}{=} \PY{n}{raw\PYZus{}to\PYZus{}chunk}\PY{p}{(}\PY{n}{p2}\PY{p}{)}\PY{p}{;}
  \PY{c+c1}{// growing size into double}
  \PY{n}{c2}\PY{o}{\PYZhy{}}\PY{o}{\PYZgt{}}\PY{n}{size} \PY{o}{=} \PY{l+m+mi}{2} \PY{o}{*} \PY{n}{chunk\PYZus{}size}\PY{p}{(}\PY{n}{sz}\PY{p}{)} \PY{o}{|} \PY{l+m+mi}{1}\PY{p}{;}

  \PY{c+c1}{// p5\PYZsq{}s chunk size = chunk\PYZus{}size(sz) * 2}
  \PY{k+kt}{void} \PY{o}{*}\PY{n}{p5} \PY{o}{=} \PY{n}{malloc}\PY{p}{(}\PY{n}{sz}\PY{p}{)}\PY{p}{;}
  \PY{c+c1}{// move p5 to the unsorted bin}
  \PY{n}{free}\PY{p}{(}\PY{n}{p5}\PY{p}{)}\PY{p}{;}

  \PY{c+c1}{// splitting p5 into half and returning p6}
  \PY{k+kt}{void}\PY{o}{*} \PY{n}{p6} \PY{o}{=} \PY{n}{malloc}\PY{p}{(}\PY{n}{sz}\PY{p}{)}\PY{p}{;}
  \PY{c+c1}{// returning the remainder}
  \PY{k+kt}{void}\PY{o}{*} \PY{n}{p7} \PY{o}{=} \PY{n}{malloc}\PY{p}{(}\PY{n}{sz}\PY{p}{)}\PY{p}{;}

  \PY{c+c1}{// p3 and p7 overlap}
  \PY{n}{assert}\PY{p}{(}\PY{n}{p3} \PY{o}{=}\PY{o}{=} \PY{n}{p7}\PY{p}{)}\PY{p}{;}
\end{Verbatim}

%% file: code/unaligned-double-free.c.tex
\begin{Verbatim}[commandchars=\\\{\},codes={\catcode`\$=3\catcode`\^=7\catcode`\_=8}]
  \PY{c+c1}{// [PRE\PYZhy{}CONDITION]}
  \PY{c+c1}{//    sz1: non\PYZhy{}fast\PYZhy{}bin size}
  \PY{c+c1}{//    sz2: non\PYZhy{}fast\PYZhy{}bin size}
  \PY{c+c1}{//    sz1 and sz2 have the following relationship;}
  \PY{c+c1}{//    assert(chunk\PYZus{}size(sz1) * a == chunk\PYZus{}size(sz2) * b);}
  \PY{c+c1}{// [BUG] double free}
  \PY{c+c1}{// [POST\PYZhy{}CONDITION]}
  \PY{c+c1}{//    two chunks overlap}

  \PY{k}{for} \PY{p}{(}\PY{k+kt}{int} \PY{n}{i} \PY{o}{=} \PY{l+m+mi}{0}\PY{p}{;} \PY{n}{i} \PY{o}{\PYZlt{}} \PY{n}{a}\PY{p}{;} \PY{n}{i}\PY{o}{+}\PY{o}{+}\PY{p}{)}
    \PY{n}{p1}\PY{p}{[}\PY{n}{i}\PY{p}{]} \PY{o}{=} \PY{n}{malloc}\PY{p}{(}\PY{n}{sz1}\PY{p}{)}\PY{p}{;}

  \PY{c+c1}{// allocate a chunk to prevent merging with the top chunk}
  \PY{k+kt}{void}\PY{o}{*} \PY{n}{p} \PY{o}{=} \PY{n}{malloc}\PY{p}{(}\PY{l+m+mi}{0}\PY{p}{)}\PY{p}{;}

  \PY{c+c1}{// free from backward not to modify size of p1[a \PYZhy{} 1]}
  \PY{k}{for} \PY{p}{(}\PY{k+kt}{int} \PY{n}{i} \PY{o}{=} \PY{n}{a} \PY{o}{\PYZhy{}} \PY{l+m+mi}{1}\PY{p}{;} \PY{n}{i} \PY{o}{\PYZgt{}}\PY{o}{=} \PY{l+m+mi}{0}\PY{p}{;} \PY{n}{i}\PY{o}{\PYZhy{}}\PY{o}{\PYZhy{}}\PY{p}{)}
    \PY{n}{free}\PY{p}{(}\PY{n}{p1}\PY{p}{[}\PY{n}{i}\PY{p}{]}\PY{p}{)}\PY{p}{;}

  \PY{c+c1}{// allocate chunks to fill empty space}
  \PY{k}{for} \PY{p}{(}\PY{k+kt}{int} \PY{n}{i} \PY{o}{=} \PY{l+m+mi}{0}\PY{p}{;} \PY{n}{i} \PY{o}{\PYZlt{}} \PY{n}{b}\PY{p}{;} \PY{n}{i}\PY{o}{+}\PY{o}{+}\PY{p}{)}
    \PY{n}{p2}\PY{p}{[}\PY{n}{i}\PY{p}{]} \PY{o}{=} \PY{n}{malloc}\PY{p}{(}\PY{n}{sz2}\PY{p}{)}\PY{p}{;}

  \PY{c+c1}{// now a next free chunk of p1[a\PYZhy{}1] is p whose P=1,}
  \PY{c+c1}{// and p1[a\PYZhy{}1] contains valid metadata}
  \PY{c+c1}{// since malloc does not clean up the memory}

  \PY{c+c1}{// [BUG] double free}
  \PY{n}{free}\PY{p}{(}\PY{n}{p1}\PY{p}{[}\PY{n}{a}\PY{o}{\PYZhy{}}\PY{l+m+mi}{1}\PY{p}{]}\PY{p}{)}\PY{p}{;}

  \PY{c+c1}{// now new allocation returns p1[a\PYZhy{}1]}
  \PY{c+c1}{// that overlaps with p2[b\PYZhy{}1]}
  \PY{n}{assert}\PY{p}{(}\PY{n}{malloc}\PY{p}{(}\PY{n}{sz1}\PY{p}{)} \PY{o}{=}\PY{o}{=} \PY{n}{p1}\PY{p}{[}\PY{n}{a}\PY{o}{\PYZhy{}}\PY{l+m+mi}{1}\PY{p}{]}\PY{p}{)}\PY{p}{;}
\end{Verbatim}

%% file: code/house-of-unsorted-einherjar.c.tex
\begin{Verbatim}[commandchars=\\\{\},codes={\catcode`\$=3\catcode`\^=7\catcode`\_=8}]
  \PY{c+c1}{// [PRE\PYZhy{}CONDITION]}
  \PY{c+c1}{//    sz: small bin size}
  \PY{c+c1}{//    assert(chunk\PYZus{}size(sz) \PYZam{} 0xff == 0);}
  \PY{c+c1}{// [BUG] off\PYZhy{}by\PYZhy{}one NULL}
  \PY{c+c1}{// [POST\PYZhy{}CONDITION]}
  \PY{c+c1}{//    assert(raw\PYZus{}to\PYZus{}chunk(malloc(sz)) == fake);}

  \PY{c+c1}{// the lowest byte of chunk\PYZus{}size(sz) needs to be zero}
  \PY{c+c1}{// to avoid chaning its size when triggering a bug}
  \PY{c+c1}{// assert(chunk\PYZus{}size(sz) \PYZam{} 0xff == 0);}
  \PY{k+kt}{char} \PY{o}{*}\PY{n}{p1} \PY{o}{=} \PY{n}{malloc}\PY{p}{(}\PY{n}{sz}\PY{p}{)}\PY{p}{;}
  \PY{k+kt}{char} \PY{o}{*}\PY{n}{p2} \PY{o}{=} \PY{n}{malloc}\PY{p}{(}\PY{n}{sz}\PY{p}{)}\PY{p}{;}
  \PY{k+kt}{char} \PY{o}{*}\PY{n}{p3} \PY{o}{=} \PY{n}{malloc}\PY{p}{(}\PY{n}{sz}\PY{p}{)}\PY{p}{;}
  \PY{k+kt}{char} \PY{o}{*}\PY{n}{p4} \PY{o}{=} \PY{n}{malloc}\PY{p}{(}\PY{n}{sz}\PY{p}{)}\PY{p}{;}

  \PY{c+c1}{// move p1 to unsorted bin}
  \PY{n}{free}\PY{p}{(}\PY{n}{p1}\PY{p}{)}\PY{p}{;}

  \PY{k}{struct} \PY{n}{malloc\PYZus{}chunk}\PY{o}{*} \PY{n}{c3} \PY{o}{=} \PY{n}{raw\PYZus{}to\PYZus{}chunk}\PY{p}{(}\PY{n}{p3}\PY{p}{)}\PY{p}{;}
  \PY{c+c1}{// make prev\PYZus{}size into double to cover a large chunk}
  \PY{c+c1}{// this is valid by writing p2\PYZsq{}s last data}
  \PY{n}{c3}\PY{o}{\PYZhy{}}\PY{o}{\PYZgt{}}\PY{n}{prev\PYZus{}size} \PY{o}{=} \PY{n}{chunk\PYZus{}size}\PY{p}{(}\PY{n}{sz}\PY{p}{)} \PY{o}{*} \PY{l+m+mi}{2}\PY{p}{;}
  \PY{c+c1}{// [BUG] use off\PYZhy{}by\PYZhy{}one NULL to make P=0 in c3}
  \PY{n}{assert}\PY{p}{(}\PY{p}{(}\PY{n}{c3}\PY{o}{\PYZhy{}}\PY{o}{\PYZgt{}}\PY{n}{size} \PY{o}{\PYZam{}} \PY{l+m+mh}{0xff}\PY{p}{)} \PY{o}{=}\PY{o}{=} \PY{l+m+mh}{0x01}\PY{p}{)}\PY{p}{;}
  \PY{n}{c3}\PY{o}{\PYZhy{}}\PY{o}{\PYZgt{}}\PY{n}{size} \PY{o}{\PYZam{}}\PY{o}{=} \PY{o}{\PYZti{}}\PY{l+m+mi}{1}\PY{p}{;}

  \PY{c+c1}{// this will merge p1 \PYZam{} p3}
  \PY{n}{free}\PY{p}{(}\PY{n}{p3}\PY{p}{)}\PY{p}{;}

  \PY{c+c1}{// if we allocate p5,}
  \PY{c+c1}{// p2 is now points to a free chunk in the unsorted bin}
  \PY{k+kt}{char} \PY{o}{*}\PY{n}{p5} \PY{o}{=} \PY{n}{malloc}\PY{p}{(}\PY{n}{sz}\PY{p}{)}\PY{p}{;}

  \PY{c+c1}{// it\PYZsq{}s unsorted bin into stack}
  \PY{k}{struct} \PY{n}{malloc\PYZus{}chunk}\PY{o}{*} \PY{n}{fake} \PY{o}{=} \PY{p}{(}\PY{k+kt}{void}\PY{o}{*}\PY{p}{)}\PY{n}{buf}\PY{p}{;}
  \PY{c+c1}{// set fake\PYZhy{}\PYZgt{}size to chunk\PYZus{}size(sz) for later allocation}
  \PY{n}{fake}\PY{o}{\PYZhy{}}\PY{o}{\PYZgt{}}\PY{n}{size} \PY{o}{=} \PY{n}{chunk\PYZus{}size}\PY{p}{(}\PY{n}{sz}\PY{p}{)}\PY{p}{;}
  \PY{c+c1}{// set fake\PYZhy{}\PYZgt{}bk to any writable address to avoid crash}
  \PY{n}{fake}\PY{o}{\PYZhy{}}\PY{o}{\PYZgt{}}\PY{n}{bk} \PY{o}{=} \PY{p}{(}\PY{k+kt}{void}\PY{o}{*}\PY{p}{)}\PY{n}{buf}\PY{p}{;}

  \PY{k}{struct} \PY{n}{malloc\PYZus{}chunk}\PY{o}{*} \PY{n}{c2} \PY{o}{=} \PY{n}{raw\PYZus{}to\PYZus{}chunk}\PY{p}{(}\PY{n}{p2}\PY{p}{)}\PY{p}{;}
  \PY{n}{c2}\PY{o}{\PYZhy{}}\PY{o}{\PYZgt{}}\PY{n}{bk} \PY{o}{=} \PY{n}{fake}\PY{p}{;}

  \PY{n}{assert}\PY{p}{(}\PY{n}{raw\PYZus{}to\PYZus{}chunk}\PY{p}{(}\PY{n}{malloc}\PY{p}{(}\PY{n}{sz}\PY{p}{)}\PY{p}{)} \PY{o}{=}\PY{o}{=} \PY{n}{fake}\PY{p}{)}\PY{p}{;}
\end{Verbatim}